\documentclass[12pt]{article}

\usepackage[latin1]{inputenc}
\usepackage[mathcal]{euscript}
\usepackage{amsmath}
\usepackage{amsfonts}
\usepackage{amssymb}
\usepackage{xcolor}
\usepackage{graphicx}
\DeclareGraphicsExtensions{.pdf, .jpg, .jpeg, .png}
\graphicspath{{.}{figures/}}
\usepackage{hyperref}
\hypersetup{ 
	colorlinks=true,
	linkcolor=blue,
	citecolor=blue,
	urlcolor=red
                   }

\usepackage{topcapt}

\newcounter{foo} 
\date{}

\begin{document}

\title{The Voronoi Diagram in Soccer\\ \large a theoretical study to measure dominance space} 
\author{Costas J. Efthimiou \\ \small Department of Physics, University of Central Florida, Orlando, FL 32816}
\maketitle

\begin{abstract}
In this article, I reexamine that the concept of the standard Voronoi diagram for the calculation of dominance area in soccer and explain why it is incorrect to use it. The correct region that should replace the polygonal Voronoi region controlled by a player  emerges naturally by studying the motion of the players. It turns out that the widely-used standard Voronoi diagram is true only when all 22 players have the same speed (with vanishing speed being a special case) and it should be replaced by other generalized Voronoi diagrams (already known to mathematicians) whose regions have more complicated boundaries.  The unfortunate `error' happened when the mathematical concept of the Voronoi diagram that has an implicit vanishing speed for its points was transferred to soccer without looking at the kinematical issues involved in the game of soccer. For this reason,   as a byproduct of this work, I  argue for the promotion of \textbf{soccerdynamics}, model building for soccer by looking at the underlying mechanisms and not just adopting mathematical ideas only for their appeal and beauty.
\end{abstract}

\newpage

\section{Introduction}

%________________________________________________________________________________________
\subsection*{\normalsize Voronoi diagrams in soccer}
Although it can be debated, the first person to deal with the concept of  Voronoi diagrams is considered to be Gustav Lejeune Dirichlet who used the two- and three-dimensional cases in his studies \cite{Dirichlet}. Quite later, Georges Voronoi studied the general $n$-dimensional case \cite{Voronoi1,Voronoi2} and, hence, the researchers who followed named the concept after him.

Simply said, Voronoi diagrams are `proximity maps'. For example, given the fire stations in a city, we can divide the city in regions such that every building of the city is placed in the the same region with its closest fire station. We can create a similar map for hospitals, schools, shopping malls, etc. Besides being used to find the closest facility of interest, the map can be used to demonstrate lack of services and to plan new actions. For example, if a hospital serves a disproportionate large region, then it can be used to decide where a new hospital must be built such as  the population is distributed more fairly.

With the growing availability of player positional data and the desire to use them in a productive way for teams, 
 it was inevitable that the concept of the Voronoi diagram would find its way in soccer. The origins seem to go back to the 1996 article of Taki, Hasegawa and Fukumura \cite{THF} and  a follow up publication in 2000 \cite{TH}.
Since then,  Voronoi diagrams were adopted by soccer (and other team sports) data analysts as the fundamental way to measure the area controlled by a player and also the entire team.
%(see, for example, \cite{Kim,Law,FS,FMTA,FMTAL,GDA}).
Besides its straightforward  use of computing control area, the Voronoi diagrams have been used in more complex ideas as a key ingredient in computing other performance indices (see, for example,\cite{SADDAG,RRPM,HGCE,CBG}).

The Voronoi diagram as a tool in soccer was popularized to a wide audience by David Sumpter in his entertaining trade book \textit{Soccermatics} \cite{Sumpter}. Sumpter started with the passing triangles of Barcelona which form, in the language of mathematics, a Delaunay triangulation. He then converted it  to a Voronoi diagram using a mathematical procedure. The two constructions provide equivalent descriptions of the underlying geometry. However, depending on the situation, one construction can be more useful than the other. In his example, Sumpter used the Voronoi diagram to visually reveal  the strength of Barcelona's great attacking movement.

One can find many blogs online which explain Voronoi diagrams for game evaluation and even blogs which describe how to draw them using one's favorite programming language (see, for example, \cite{web1,web2,web3}).

%_______________________________________________________________________
\subsection*{\normalsize What the Voronoi diagrams are and what they are not}

A reader with a penetrating understanding has already understood from the previous introduction that the Voronoi diagrams can be used
for at least two distinct concepts in soccer:
Sumpter's use of the Voronoi diagrams as a \textit{a measure for attacking strength} and the standard use of 
%does not automatically explain the use of 
Voronoi diagrams as  \textit{a measure of dominance area}. 
%These are two different concepts: attacking strength of the passing triangles and team dominance area. 
The two concepts are not equivalent since they evaluate different aspects of a team. This becomes immediately evident if you think of the following two issues: (a) It is possible that a team dominates most of the pitch since its players stand further away from the action, but the other team has the best passing triangles in the tight space. (b) To evaluate the quality of the passing triangles, we draw the Voronoi diagram for the 11 players of the same team. However, to evaluate which team controls more space, we draw the Voronoi diagram for the totality of the 22 players in the pitch.  Having explained the above, in this article I will focus on the Voronoi diagrams as a measure of dominance space.

Researchers  (including myself) criticize or defend the Voronoi diagrams based on their personal preferences. One property that one researcher might see as an advantage, another researcher might see it as a disadvantage. Therefore, exactly as the title suggests, I would like to use this subsection to present my arguments what the Voronoi diagrams are and what they are not.

Based on its definition, the Voronoi diagram is a set of the proximity regions of the players. Given the location of the players P$_i$, $i=1,2,\dots,22$ at a fixed time $T$, the pitch is partitioned in regions V$_i$, $i=1,2,\dots,22$ such that V$_i$ contains the points of the pitch
which are nearer to P$_i$ than any of the remaining players. This mathematical proximity statement (which does not include time), when applied to soccer,  is equivalent to the following process (which contains time-dependent actions): 
At each point P$_i$, we imagine an infinite number of identical players, one for each ray emanating from the point P$_i$.
We allow all players to start running simultaneously with zero initial speed. After  time $\Delta t$, 
we mark the points reached by the copies of the player P$_i$. Among them, those points which cannot be reached by any copy of another player at a lesser time  are declared as  points of the dominance region V$_i$. We repeat this process for all possible values of $\Delta t$
and at the end the regions V$_i$ have been found.  
Hence, the mathematician's  Voronoi diagram is an abstract topological construction which has nothing to do with time but
the soccer analyst's Voronoi diagram is a procedural concept, a diagram that can be created only after a particular procedure involving time is executed. 
As an extrapolation of this comment, the reader can realize immediately two important properties: 
(a)  The regions V$_i$ are the \textit{instantaneous} dominance regions of the players at the time $T$. 
That is, given the positions $\vec r_i$ of all players, the regions $V_i$ are functions of them. Moreover, if we know the trajectories of
the players, $\vec r_i= \vec r_i(t)$, then the regions $V_i$ should vary \textit{continuously} on them. And, since the trajectories of the players are continuous functions of time, the dominance regions must be so too:  $V_i=V_i(\vec r_1(t),\dots, \vec r_{22}(t))=\tilde V_i(t)$. Notice, however, that the regions $V_i$ have no explicit dependence on time; time enters implicitly from the motion of the players.\footnote{This assumption can easily be modified: One can add a fatigue parameter for the players. As time passes, the fatigue parameter becomes more important for the
way players would run to compete for space. In this case, the construction of the dominance regions is time-dependent and the dominance
regions $V_i$, besides the implicit dependence on time, they acquire also an explicit  time-dependence.}
(b) The Voronoi diagram at any time $T$ \textit{cannot} be constructed solely from the tracking data of the players.
Drawing the boundaries of the dominance regions requires additional assumptions about the way the players \textit{would run}, which cannot be derived by the real-time data. The standard Voronoi diagram assumes that this hypothetical running happens such that all players start  from rest and move uniformly with the same speed along any direction. Since these rules for the construction of the dominance regions are $T$-independent, this makes the regions not to be explicit functions of time as indicated previously.

The dominance regions of the Voronoi diagram are \textit{not} ball control regions --- that is, regions where a player will necessarily intercept the ball first. For such a diagram, additional parameters  must necessarily be included since there is an additional object of interest.

Similarly to the concepts of expected goals, expected assists and expected threat, the Voronoi diagram 
is a theoretical construction that attempts to provide useful insight for the game.  However, contrary to the `expected quantities' which are probabilistic statements quantifying possibilities that \textit{may eventually happen} using events which have happened, the Voronoi diagram is a deterministic construction that quantifies  the movement of players and teams as it is \textit{actually happening}. 

The standard Voronoi diagram is simple and straightforward but inadequate. Improved versions 
can be modeled using well understood physical laws and popular mathematical tools.  The modeling can be further improved by understanding better the physiological principles of the hypothetical running which determines the 
dominance regions.
%As such, it can be improved as our understanding of the game improves.
%
%The use of the Voronoi diagram assumes that things happen in an optimal way. For example, a player who can reach a point first, he will indeed reach it first. 
%That is, no mistake is involved. In \cite{Fernandez1}, the authors state: ``This concept disregards the concept that ownership of space is continuous, not discrete, with uncertainty in who controls areas between players."  Actually, ownership is discrete. Under ideal circumstances, there is a single player that can reach a point first. 
%
%XXXX MAYBE IT GOES TO THE NEXT SECTION
%As I will demonstrate explicitly  in the next section, \textit{the straightforward application of Voronoi diagrams in soccer as a measure for dominance area without taking into account the reasoning that leads to them is flawed}. Pure mathematical modeling is unfortunately a superficial way to explain natural phenomena since it makes implicit assumptions about reality that may not be correct when one looks at the mechanism that underlies the phenomenon. Said in another way, mathematics is only a tool, a `device' that allows us to reason correctly; its statements cannot be accepted as physical laws without further investigation. 
Hence, besides \textit{soccermatics} (application of math to soccer), 
I want to advance the concept  of \textbf{soccerdynamics}, that is, the understanding of the underlying dynamics (mechanism)  at work in any soccer situation.  It is easily understood that this goal is often plagued with insurmountable difficulty which increases as one requires a finer  level of understanding.

%_________________________________________________________
\subsection*{\normalsize Comparing this work with other similar works}

In this work, I construct specific models for the dominance areas which take into account not only the position of the players but also the speeds at the specific moment of time. In addition, I introduce two new parameters: the maximal speed a player can reach and his reaction time. The computational details of the models are based on simple kinematics and well known mathematical tools.  They have explicit analytical solutions which  provide good insights directly related to the soccer game. 

Since the introduction of the dominance regions\footnote{It is of interest to note that the term `Voronoi regions' is not present in the original article \cite{THF}; it first appeared in the follow up publication \cite{TH}.}  to soccer analytics, there have been various efforts to improve the concept. In the remaining of this section, I will add some comments on those past efforts and how they compare to the work presented in the article.

In the original papers, Taki and collaborators \cite{THF,TH} were working with video cameras which was making their work to extract the dominance regions quite laborious. Since these original papers, two main computational advances occurred.  In 2004, Kim bypassed the  problem of having explicit tracking data by using data borrowed from the EA Sports' FIFA 2003 game \cite{Kim}. Kim provides no explicit reasoning  or motivation why the standard Voronoi diagrams are used for the match analysis; nor he refers to \cite{THF,TH}.  It appears that the property of Voronoi diagrams of being `proximity maps' is silently adopted as the qualifying one. The paper makes no attempt to evaluate how well the concept fits in the soccer environment.  However, with Kim's paper providing  an example how to draw Voronoi diagrams for soccer games and tracking data for players starting to be available, the Voronoi diagrams started to increase in popularity due to its conceptual simplicity and the ease they could be drawn. In the  RoboCup International Symposium 2009, motivated by the original papers \cite{THF,TH}, Nakanishi et al. proposed an approximate algorithm for a real-time calculation of the dominance area  which shortens the calculation time by a factor of $10^{3}$ \cite{NMMN}. This  is a great  advance  for computational work, but it does not provide any new insight from an analytical perspective.

With the increasing popularity of the standard Voronoi diagrams, the voices warning about the limited usefulness of the diagrams because they do not incorporate the players' motion increased too \cite{BCF}.  Moreover,  in all papers that build probabilistic models for game actions and performance indices the dominance regions is introduced and used. Claiming the insufficiency the standard Voronoi diagrams, these papers often propose an improved version for the computation of the dominance regions but, unfortunately,  only ad hoc constructions are proposed, not based on careful investigations and detailed explanations. For example: In \cite{BCF,CBG}, a modified Voronoi diagram is introduced through an intuitive analogy. In particular, the degree of ownership is assigned a weight which is inversely proportional to the distance. However, no attempt is made to reason for the selection of the weight using arguments based on the game.  In \cite{Fernandez1}, a player influence area is defined as a multivariate normal distribution. In this proposal, besides the player's location, the player's speed and the ball's location is taken into account.  The model is pulled out of the hat without any kinematical argument driven by actual game dynamics. 

An interesting paper in agreement with my philosophy and expectations (for deterministic models motivated by physics) has been presented by Burris \cite{Burris}. Although it was written for the American football, its content can be relevant to soccer. In this paper, using results from robotics and autonomous vehicles in constrained non-linear programming, the author discusses optimal player trajectory planning. The author is not interested so much in Voronoi diagrams but
in players moving to the optimal position to receive a successful passing. In his own words:
%\footnote{Unfortunately, it appears to be a misconception here. Pure Voronoi diagrams alone cannot predict successful passing. He may be comparing his results to probabilistic models of passing but there is no cited reference and data to validate the statement.}
 ``..., we are able to identify when a receiver beats a defender well before Voronoi-based models."

There have been many more papers that discuss pitch control and effective passing from many different points of view, some  of  which do not necessarily contain explicitly the Voronoi diagrams as an ingredient  (e.g. \cite{Spearman1, Spearman2, ASMFHB, MDB, Wal}). What all these papers
have in common are (a) a set of arbitrarily chosen assumptions, (b)  a probabilistic model for pitch control and (c) the ball is included in the model. However, as I have mentioned, the Voronoi diagram is \textit{not} a ball control diagram.
The latter is a higher level concept. A model for the ball control diagram can be built with or without the use of the Voronoi diagram. Such a model is not my goal is this article; I will stay confined within the limits of the dominance 
regions. Also, in my mind, being able to reach a point first is  both binary (the player can or cannot do it) and deterministic (the result can be worked out out with absolute certainty given the initial conditions). Hence, I believe that probabilistic 
models are completely unnecessary for this particular concept and they become more important for ball control. To the best of my knowledge, there is  no dedicated theoretical work that reviews critically the dominance regions concept and 
builds in-depth analytical models based on simple, reasonable kinematical assumptions. It is  this
vacuum in the literature that I wish to fill.

%%%%%%%%%%%%%%%%%%%%%%%%%%%%%%%%%%%%%%%%%%%%%%%%%%
\section{Dominance area}

Given a set of points P$_i$, $i=1,2,\dots,22$, on the plane $\mathbb{R}^2$, the Voronoi region of the point P$_i$ is the set of points
$$
         \overline V_i = \{ \text{P}\in\mathbb{R}^2 ~|~  d(\text{P}, \text{P}_i) \le d(\text{P},\text{P}_j) ~\forall j\ne i\},
$$
where $d(\text{A, B})$ denotes the Euclidean distance between the points A and B. The set of of points
$$
          V_i = \{ P\in\mathbb{R}^2 ~|~  d(\text{P}, \text{P}_i) < d(\text{P},\text{P}_j) ~\forall j\ne i\},
$$
is the called the open Voronoi region and the set
$$
    \partial V_i = \overline V_i \smallsetminus V_i
$$
is the boundary of the Voronoi region $V_i$.  Since soccer is played in a closed domain $S$ of $\mathbb{R}^2$, we are interested in  the 
 intersections $S\cap V_i$. Obviously, only those regions through which the boundary $\partial S$  of $S$ passes  are modified. 
 We draw a Voronoi diagram by drawing the boundaries $\partial V_i$ for all players P$_i$.

The above is the definition that is used currently  in soccer when Voronoi diagrams are used to compute dominance area of a team. 
However, it is quite obvious that the above definition implies that the points (players) are at rest. Hence, computing the Voronoi diagram at kick-off (or at any other instant at which all 22 players are at rest) will give a correct result of how much area each team dominates at that moment. 
\begin{figure}[h!]
\centering
\includegraphics[width=8cm]{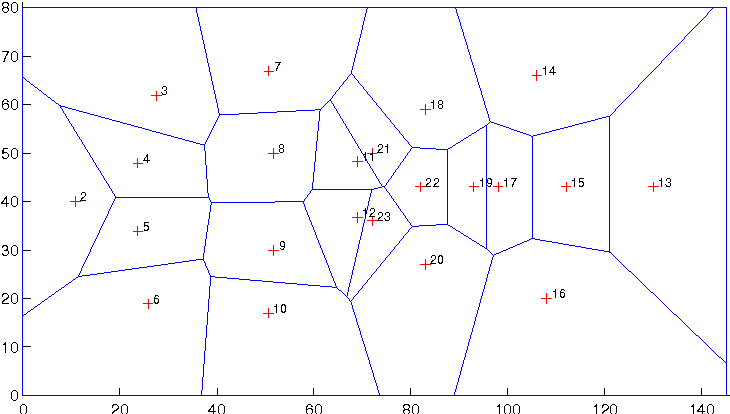}
\caption{\footnotesize A typical Voronoi diagram for the 22 players (labelled 2--23). This diagram is taken from \cite{Kim}. }
\label{fig:Voronoi}
\end{figure}
On the other hand, there is no reason to expect that this definition provides a faithful calculation for the dominance area at all remaining times since not all players are simultaneously at rest. Soccer is a game where motion is of fundamental importance and, as such, it cannot be ignored. Therefore, it is already evident that the standard Voronoi diagrams can  only  be a rough estimate; they may actually miscalculate the dominance area. But without a model of dominance area that is based on the players' motion, it is impossible to judge the degree of precision of the Voronoi diagrams.

Mathematics creates concepts but how the concepts are applied are upon to the people who create the models. And models are as good as the assumptions used to create them.  In the following, I improve the model by introducing the motion of players. Of course, some assumptions are made to keep the models simple. These assumptions can certainly be modified and make the models as sophisticated as one may like.  My intension is to initiate a discussion  for improving the ideas currently used to compute dominance area of a player and a team.

%___________________________
\subsection*{\normalsize Uniform motion}
Consider two players P$_1$ and P$_2$ at rest and let $r_1, r_2$ be their respective  distances from any point P on the plane. 
When the players try to reach the point P, we assume they do so with uniform speeds $V_1$ and $V_2$ respectively which they can acquire instantly. If they start at the same time, say $t=0$,  and they both reach the  point P  at time $t$,
\begin{equation}
   r_1=V_1 \, t, \quad r_2=V_2\, t.
\label{eq:1}
\end{equation}
Therefore, the points that the two players can reach simultaneously satisfy the equation
\begin{equation}
     {r_1 \over r_2} = {V_1 \over V_2} = \text{const}.
\label{eq:5}
\end{equation}

Equations like equation \eqref{eq:5} which use two points as reference points are known as \textbf{bipolar equations}. The two reference points are called 
\textbf{foci}. In particular, the bipolar equation
$$
     {r_1\over r_2} = \lambda,
$$
with $\lambda>0$  is probably the most widely known from Euclidean geometry: All points that satisfy it lie on a circle which is known as the \textbf{Apollonius circle} named after the great geometer Apollonius of Perga who was the first person to study this problem. It will be constructed explicitly in Section \ref{sec:AC}.

We have thus found that the boundary $\partial V$ between two players is a circle. If $V_1=V_2$, then $r_1=r_2$ and the circle degenerates to the perpendicular
bisector of the segment P$_1$P$_2$.
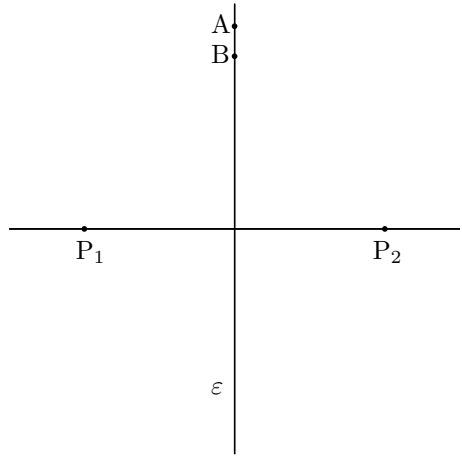
\begin{figure}[h!]
\centering
\setlength{\unitlength}{1mm}
\begin{picture}(60,55)
\put(0,-3){\includegraphics[width=6cm]{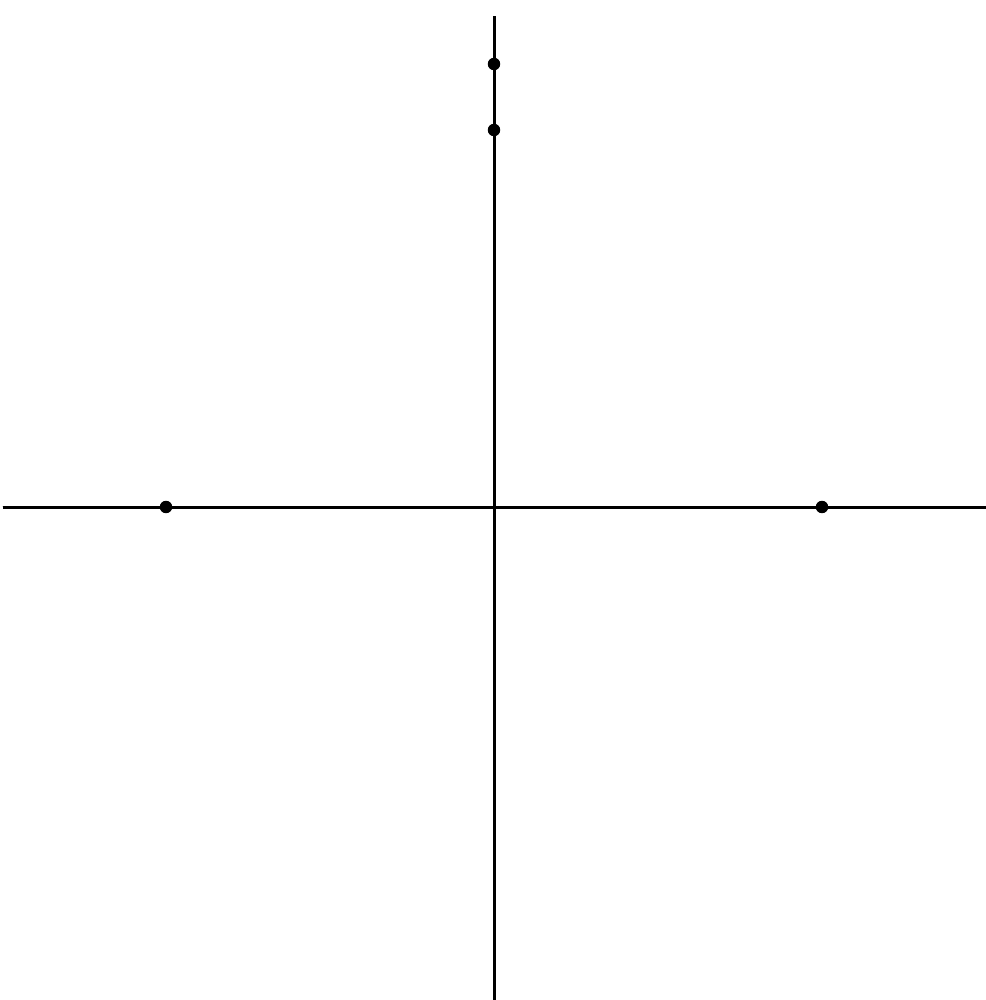}}
\put(9,23){\footnotesize P$_1$}
\put(48.5,23){\footnotesize P$_2$}
\put(27,53){\footnotesize A}
\put(27,49){\footnotesize B}
\put(27,5){\footnotesize $\varepsilon$}
\end{picture}
\caption{\footnotesize The boundary of the dominance regions, when the players P$_1$, P$_2$ have equal speeds 
             or are at rest  is the perpendicular bisector $\varepsilon$ of P$_1$P$_2$. Notice that, although $\varepsilon$ contains points which are reached simultaneously
             by the players,  different points on $\varepsilon$, such as A and B, 
             require different times to be reached. The same is of course true for points on the Apollonius circle.}
\label{fig:bisector}
\end{figure}
%

%_____________________________________________________________
\subsection*{\normalsize Accelerated motion I}
The previous calculation has some shortcomings that we will resolve by adopting improved assumptions. In particular, when players attempt to run they cannot start immediately with their final velocity. That is, their motion cannot be uniform; it has to be accelerated. We will assume a uniform acceleration for simplicity. Let it be $a_1$ and $a_2$ respectively for the two players. Therefore, equations  \eqref{eq:1} for the distances $r_1, r_2$ must be modified to
\begin{equation}
   r_1= {1\over2} \, a_1 \, t^2, \quad r_2= {1\over2} \, a_2 \, t^2 .
\label{eq:2}
\end{equation}
The points that the two players can reach simultaneously satisfy now the equation
$$
     {r_1 \over r_2} = {a_1 \over a_2},
$$
which is, again, an  Apollonius circle.
% if we assume that the players always use the same value of acceleration.

The previous result faces two objections.  First it appears that the circle is now defined  by the accelerations of the players and not their speeds. 
This objection, of course, is not a serious one. The two motions are different, so the new scenario is not obligated to lead to identical results. However, the second objection is more substantial: From experience, we know that players do not accelerate  identically along all directions. 
We can actually resolve both issues with a slight modification in our assumptions.

The two quantities, speed and acceleration, are ultimately related. In accelerated motion with zero initial speed, the final speed is proportional to the acceleration; therefore,  $V_i=a_i \, t$, $i=1,2$ and hence we recover a similar result to equation \eqref{eq:5}:
$$
     {r_1 \over r_2} = {V_1 \over V_2} .
$$
%But now a new problem has appeared. 
For different points P of the circle, the ratio of speeds will have a different value if the acceleration of the players along different directions has the same value. To overcome this difficulty, I will assume that the players reach the common point P with their maximum speed. In other words, I assume that, as players accelerate in each direction, they can set the magnitude of the acceleration  to whatever value is necessary for this to happen. The final speeds are now specific constants for  the players and the ratio $V_1/V_2$ is constant. 
As a reminder that the speed $V_i$ is a  fixed constant for player P$_i$, I will name it the \textbf{characteristic speed} of player P$_i$. 
It is important to understand that, in the modified approach, the ratio of accelerations does not have a constant ratio anymore. We cannot have
both ratios $a_1/a_2$, $V_1/V_2$ constant; a choice must be made and the latter choice fits soccer better.

%__________________________________________________
\subsection*{\normalsize Accelerated motion II}
In the previous accelerated model, we assumed that the players were starting from rest. However, in soccer players are always in transitive motion; hence, they have some initial velocity. So, let $\vec v_1, \vec v_2$ be the initial velocities and $\vec V_1, \vec V_2$ be the final velocities when they reach the point P.  
Notice that I am now using vectors instead of magnitudes since the direction of the various quantities do not have to coincide.
Also let's assume that each player reacts at different time to make a run towards the point P. Player one reacts at time $t_1$ and player two at time $t_2$. Besides actual delays in the response of the players, the delay time can be used to parametrize additional actions that may be necessary before the player starts accelerating. For example, if the ball is played behind the player, he will need to perform one 180$^\circ$-rotation about his axis first before he accelerates.  

Hence at time $t$, the location of each player will be given by
$$
    \vec r_i = \vec v_i \, (t-t_i) + {1\over 2} \, \vec a_i \, (t-t_i)^2, \quad i=1,2.
$$
Similarly, the velocities of the two players will be
$$
    \vec V_i = \vec v_i + \vec a_i \, (t-t_i) , \quad i=1,2.
$$
Writing  the previous equations, I have assumed that the acceleration is uniform, i.e. it has constant magnitude and direction. However, other than keeping it constant,
each player can decide to orient his acceleration in any way he likes relative to his initial velocity and adjust its magnitude such that at the point P, the velocity $\vec V_i$ 
has magnitude equal to the characteristic speed.

Although it may not completely evident, the appearance of vectors in the equations has complicated the problem considerably.
To simplify it, I will make use of the delay time $t_i$.  I will assume that the player takes a little time to reorient completely his initial speed along the
direction of the point P. In this case, the acceleration has also to be aligned along the same line and the vectorial equations become algebraic.  
With the help of the second relation, we can rewrite the first in the form
\begin{equation}
     r_i =   A_i \, (t-t_i), \quad i=1,2,
 \label{eq:15}
\end{equation}
where
$$
     A_i = {  v_i +  V_i \over 2 } , \quad i=1,2,
$$
is the average of the initial and characteristic speeds.  Then
\begin{equation}
     r_i =   A_i \, (t-t_i), \quad i=1,2.
\label{eq:4}
\end{equation}
So, the points that the two players can reach simltaneously satisfy
\begin{equation}
     {r_1 \over r_2} = {A_1 \, (t-t_1) \over A_2\, (t-t_2)} .
\label{eq:6}
\end{equation}
Incidentally, notice that this result includes the uniform motion as a special case.
% where $a_i$ for one or both players.  

We will return to the general case when each player has a different reaction time; at the moment we will assume that $t_1=t_2$. Professional players have been trained long hours to respond instinctively to the game.  That is, they routinely orient their bodies to the optimal  position ready to accelerate and their reaction times are as low as possible. Hence, to a very good approximation for the majority of the instances, we can  set $t_1=t_2$, thus recovering an Apollonius circle:
\begin{equation}
     {r_1 \over r_2} = {A_1  \over A_2} = \text{const.} ,
\label{eq:3}
\end{equation}
for the point which the can reach simultaneously.

Therefore, it is immediate from the above elementary calculations  that the standard way of
computing dominance area using Voronoi diagrams with polygonal regions is not appropriate.  Instead, a diagram based on  Apollonius circles provides
 a more faithful representation.

%%%%%%%%%%%%%%%%%%%%%%%%%%%%%%%%%%%%%%%%%%%%%%%%%%%%%
\section{Properties of the Apollonius Circle}
\label{sec:AC}

In this section, I will review quickly the properties of the Apollonius circle. In doing so, I will use analytic geometry and not theoretical geometry since the former is by far easier for practical applications (such as soccer analytics). 

To study the equation
\begin{equation}
       {r_1\over r_2} = \lambda,
\label{eq:AC0}
\end{equation}
let's use  a coordinated system O$xy$ such that the two players P$_1$ and P$_2$ are positioned at the points $(-c,0)$ and $(+c,0)$ respexctively. Without loss of generality we can assume that $\lambda>1$ since,  if this is not the case, we can relabel the players $1\leftrightarrow2$  such that it is true.

For $\lambda=1$ (which implies the two players have equal  speeds $A_1=A_2$), equation \eqref{eq:AC0} is the locus of points that satisfy $r_1=r_2$. These are the points 
of the perpendicular bisector $x=0$ of the segment that joins the two players. Therefore, the Apollonius region of each player is a half plane with boundary the 
$y$-axis, thus coinciding with the standard Voronoi regions of the players.

For $\lambda\ne1$ (which implies that $A_1>A_2$), we have to work a little harder. Given a point $(x,y)$ in the plane,
$r_1^2= (x+c)^2+y^2$ and $r_2^2=(x-c)^2+y^2$. If the point satisfies the equation \eqref{eq:AC0}, then
$$
          (x+c)^2+y^2  =  \lambda^2 \, [(x-c)^2+y^2] .
$$
After a few algebraic manipulations, we rewrite it as
\begin{equation}
      \left( x - c\,{\lambda^2+1\over \lambda^2-1} \right)^2 + y^2 = \left( c\, {2\lambda\over \lambda^2-1} \right)^2 .
 \label{eq:AC1}
 \end{equation}
This is the equation of a circle with center $(x_0, y_0)$  at
$$
    \left(  c\,{\lambda^2+1\over \lambda^2-1}  ~,~   0 \right)
$$
and radius
$$
  R = c \, {2\lambda\over  \lambda^2-1 }.
$$
If we define the quantity $\beta$ such that $\lambda=e^\beta$, we can write
$$
     x_0 = c \, \coth\beta, \quad R= {c\over \sinh\beta} .
$$
It is often advantageous for calculations to use hyperbolic trigonometry.
\begin{figure}[h!]
\centering
\includegraphics[width=10cm]{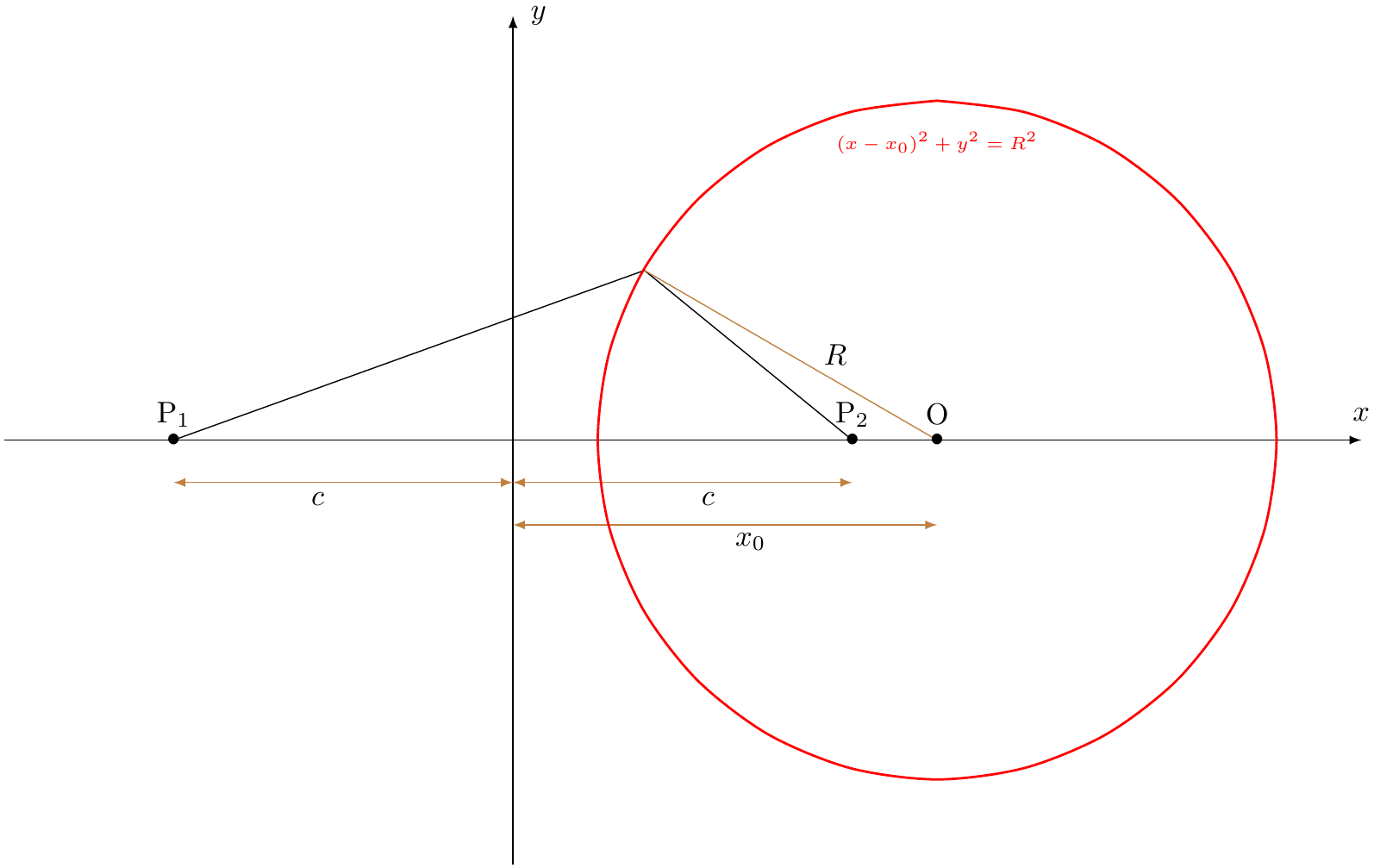}
\caption{\footnotesize Dominance areas for the two players P$_1$ and P$_2$ having  speeds $A_1>A_2$ and same reaction times with $t_1=t_2$. 
              The slower player P$_2$ controls the area inside the circle, while player P$_1$ controls the remaining area of the plane.}
\label{fig:circle}
\end{figure}

The Apollonius circle has its center always on the line that joins the two players. The abscissa of the center can be rewritten as
$$
    x_0 =   c+ {R\over \lambda} .
$$
That is, the center is located to the right of the slower player at a distance less that $R$. Hence, the Apollonius circle encloses encloses the slower player. 
(The player is inside his dominance region as expected.)
The faster the fast player (i.e. the bigger the $\lambda$), the closer the center is to the slower player and the smaller the radius of the circle. 
The slower the fast player relative to the slow player (i.e. the closer $\lambda$ is to 1), the further the center of the Apollonius circle from the slow player and the greater the
radius.
In fact, the $\lambda=1$ case can be recovered by the $\lambda\ne1$ one. In the limit $\lambda\to 1^+$,  the curvature of the circle $\kappa=1/R$ vanishes:
$$
     \kappa = {1\over R} = {1\over c} \, {\lambda^2-1 \over 2\lambda} \to 0 .
$$
This implies that the Apollonius circle degenerates to a line. This line is the $y$-axis which we can verify quickly as follows. The leftmost point of the Apollonius circle has an abscissa
$$
    x_0 - R = R \, (\cosh\beta - 1) .
$$
As $\lambda\to1$, we have $\beta\to0$ and hence $x_0-R=0$.

Notice the important difference from the standard case: The Apollonius region of the slower player is the finite region 
contained by the Apollonius circle, while the faster player dominates the remaining infinite complement. In the original scenario, we would have claimed that the difference in the dominance area is zero (each player controls a half-plane)  but in the new scenario, the faster player has an infinite 
surplus.\footnote{Notice two important assumptions for the infinite plane case: No matter how long the distance of a point is from the players: (a) they have the endurance to run the entire length and (b) they still have the ability to perform uniformly accelerated motion. Experience though implies that, for very long distances, the motion will be very different from uniformly accelerated motion. The players initially may accelerate fast, then they slow down, then they may accelerate again and eventually they may cruise at constant speed, lower than their characteristic speed, towards the destination point. Hence at least one new parameter must enter the analysis quantifying the endurance of the players  and the motion must be allowed to be more complicated. Fortunately, the finite size of the soccer pitch allows us to stay confined within a simpler model.} 
Of course, with the finite area of the soccer field, the surplus is finite  but the conceptual change that has occurred is certainly striking.

%%%%%%%%%%%%%%%%%%%%%%%%%%%%%%%%%%%%%%%%%%%%%%%%%%%%
\section{The Apollonius Diagram and its Properties}

Given the results of the previous section, we have now established a new requirement to define proximity regions in soccer. Let's define them formally:
Given a set of points P$_i$, $i=1,2,\dots,22$, on the plane $\mathbb{R}^2$, the dominance region of the point P$_i$ is the set of points
$$
         \overline V_i = \{ \text{P} \in\mathbb{R}^2 ~|~  {d(\text{P, P}_i) \over d(\text{P, P}_j)} \le  c_{ij} ,  ~\forall j\ne i\},
$$
where $c_{ij}$ is a constant matrix with the property $c_{ij}=1/c_{ji}$. Knowing that the boundary between any two points will be, in general,  
an Apollonius circles,  we may call this dominance area the \textbf{Apollonius region}. Similarly, the set of of points
$$
          V_i = \{ \text{P} \in\mathbb{R}^2 ~|~     {d(\text{P}, \text{P}_i) \over d(\text{P},\text{P}_j)} <  c_{ij} ,       ~\forall j\ne i\},
$$
will be  called the \textbf{open Apollonius region} and the set
$$
    \partial V_i = \overline V_i \smallsetminus V_i
$$
will the boundary of the Apollonius region $V_i$.  If  $S$ is the domain of $\mathbb{R}^2$ representing the field, we are interested in  the 
in the intersections $S\cap V_i$. The diagram that contains all Apollonius regions for the 22 players will be called the \textbf{Apollonius diagram}.
%If we draw all Apollonius circles for the 22 players, we get a diagram that I will call Apollonius diagram. 

It turns out that mathematicians have already proposed this variation of the Voronoi diagram although its literature is not as extensive as the standard diagram.  It has been called the \textbf{multiplicatively weighted Voronoi diagram} \cite{AE,OkabeEtAl}. Since it is one of the many possible variations
that mathematicians have invented using abstract logic,  from the strict mathematical definition, it is not easy to
realize that this particular variation is the necessary concept to adopt in soccer. Only when we study  the mechanism  that creates the natural borders of the regions of dominance, it emerges naturally. 
\begin{figure}[h!]
\centering
\includegraphics[width=8cm]{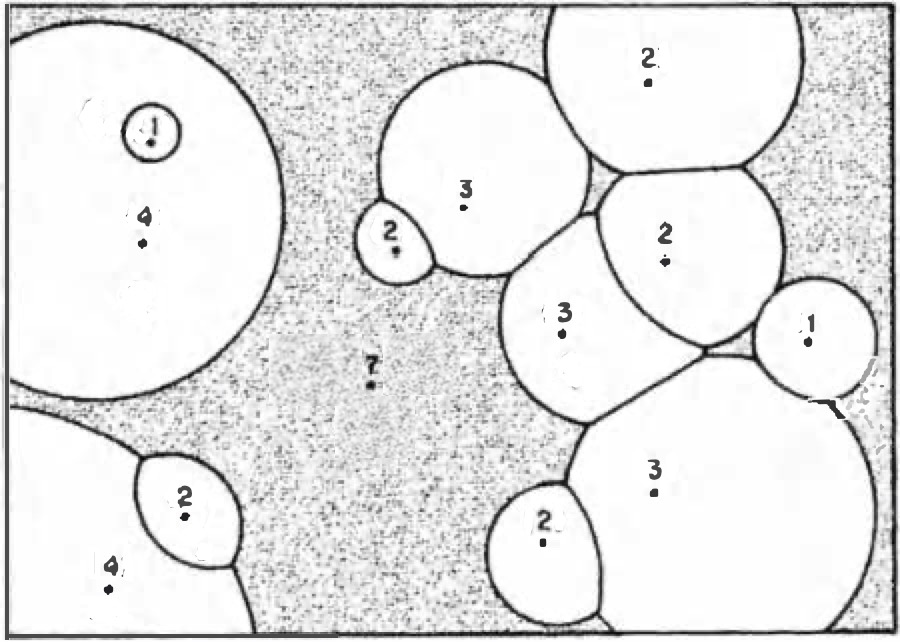}
\caption{\footnotesize An Apollonius diagram of 13 players. The numbers next to each player indicate the corresponding speed $A$. They are not necessarily realist numbers but are convenient to reveal the properties of the diagram.
This diagram is taken from \cite{OkabeEtAl}. }
\label{fig:WV1}
\end{figure}

I will list properties of the Apollonius diagram without presenting their proofs. These properties are actually evident from what has been discussed so
far  and  present in the sample Apollonius diagram shown in Figure \ref{fig:WV1}. A demanding reader can prove them easily.
\begin{enumerate}
\item An Apollonius region does not have to be convex.\footnote{Convexity is the mathematical property that requires that, given two points in a set, all
        points of the segment that joins them also belong to the set.} However, it will be convex if all adjacent regions belong to players with speeds not
        smaller than the speed of the player who owns the region.
\item An Apollonius region may contain holes. 
\item An Apollonius region may be disconnected.
\item The fastest player may dominate points far away from the rest players. This is more evident when the players have moved on one side of the field.
\item The borders between regions are either circular arcs (when the speeds of the players owning the regions are unequal) 
          or straight segments (when the  speeds of the players are equal). The border
         between  two regions may be consisting of disconnected pieces. 
\end{enumerate}

%%%%%%%%%%%%%%%%%%%%%%%%%%%%%%%%%%%%%%%%%%%%%%%%%%%%
\section{Taking into Account the Reaction Times}

\subsection*{\normalsize Equal speeds}
Let's return to equations \eqref{eq:4}. If the two players have the same speed --- say $A$ --- but different reaction times, then
\begin{equation}
        r_1 -r_2 =  A\, t_0,
\label{eq:RT0}
\end{equation}
where $t_0=t_2 - t_1$. Without loss of generality we assume that the second player has a higher delay $t_2>t_1$ (otherwise, relabel the players). 
The right-hand side of this equation is constant. Hence, we recognize equation \eqref{eq:RT0} as  a single-branch hyperbola (only those points with $r_1>r_2$) with foci at the 
locations of the two players.   

As an interesting curiosity, we set $k=1/A$ and rewrite  equation \eqref{eq:RT0} in the form
\begin{equation}
        k(r_1 -r_2) =   t_0 .
\label{eq:RT1}
\end{equation}
The left-hand side resembles now the phase difference of two sinusoidal waves with wavenumber $k$ when they arrive a point. Hence, this equation determines the interference patterns of the two waves originating at the location of the two players. Instead of wavenumber, we will call the quantity $k$ in our context the
\textbf{slowness} of the player.

\begin{figure}[h!]
\centering
\includegraphics[width=8cm]{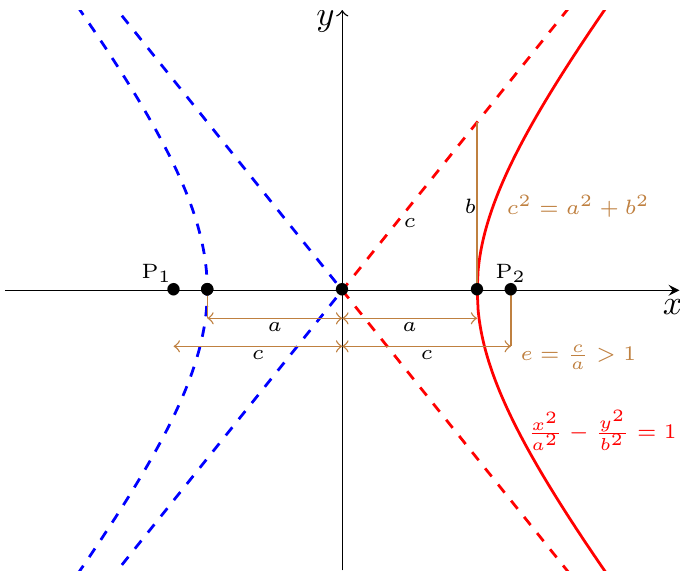}
\caption{\footnotesize Dominance areas for the two players P$_1$ and P$_2$ having the same speed $A$ but different reaction times with $t_2>t_1$. 
%Player P$_1$ controls the halfplane that contains the origin while player P$_2$ contains its complement. The two halfplanes are separated by a hyperbola (shown in red). 
The figure also shows the most important properties of the hyperbola.
%Without the reaction times, each player would control a halfplane whose boundary is the $y$-axis. Hence, the faster reaction of player P$_1$ allows him to claim additional space from player P$_2$ --- all the area between the $y$-axis and the hyperbola.
}
\label{fig:hyperbola}
\end{figure}

Using the same coordinate system we used to work out equation \eqref{eq:AC0} --- that is, the players located at $(-c,0)$ and $(c,0)$ --- and similar algebra,
we find the equation for the hyperbola to be
$$
   {x^2 \over a^2} -  {y^2\over b^2} =  1, 
$$
where
$$
      a = {At_0\over2}, \quad
      b^2=  c^2 -a^2 .
$$
This hyperbola is drawn in Figure \ref{fig:hyperbola} where the relevant branch is shown as a solid line.
Player P$_1$ controls the `curved' halfplane that contains the origin while player P$_2$ controls its complement.
Without taking into account the reaction times, each player would control a halfplane whose boundary is the $y$-axis. Hence, the faster reaction of player P$_1$ allows him to claim additional space from player P$_2$ --- all the area between the $y$-axis and the relevant branch of the hyperbola.

The greater the difference in the reaction times, the greater the value of $a$ and the smaller the value of $b$.
Hence, the closer the vertex of the hyperbola to the player P$_2$ and the closer the asymptotes move to the $x$-axis, thus reducing player's P$_2$ area.
When the delay is such that $a=c$, then the dominance area of P$_2$ is not area at all! it has been reduced to the part of the $x$-axis  that starts at his point and goes to infinity. For a greater delay than this, there is no hyperbola.  P$_2$ controls only his point!   Incidentally, notice the idealization that it is implicit in any Voronoi diagram calculation and it is so clearly revealed here: The players are considered mathematical points without dimensions. As such, one point does not create an obstruction to the other point. But players do have dimensions. This implies that their physical dimensions have also implications for controlling an area. No matter how slow a player reacts, he can certainly claim more than a single point!

Given a set of points P$_i$, $i=1,2,\dots,22$, on the plane $\mathbb{R}^2$, the dominance region of the point P$_i$ is the set of points
$$
         \overline V_i = \{ \text{P} \in\mathbb{R}^2 ~|~  {d(\text{P, P}_i) - d(\text{P, P}_j)} \le  c_{ij} ,  ~\forall j\ne i\},
$$
where $c_{ij}$ is a constant matrix with the property $c_{ij}=-c_{ji}$. Knowing that the boundary between any two points will be, in general,  hyperbolic,  we may call this dominance area the \textbf{hyperbolic region}. Similarly, the set of of points
$$
          V_i = \{ \text{P} \in\mathbb{R}^2 ~|~     {d(\text{P}, \text{P}_i) - d(\text{P},\text{P}_j)} <  c_{ij} ,       ~\forall j\ne i\},
$$
will be  called the \textbf{open hyperbolic region} and the set
$$
    \partial V_i = \overline V_i \smallsetminus V_i
$$
will the boundary of the hyperbolic region $V_i$.  If  $S$ is the domain of $\mathbb{R}^2$ representing the field, we are interested in  the 
in the intersections $S\cap V_i$. 
The diagram that contains all hyperbolic regions for the 22 players will be called the \textbf{hyperbolic diagram}.

As before, I will list some evident properties of the  diagram. 
\begin{enumerate}
\item A hyperbolic region can be empty. Even when it is non-empty, it can be a half-line or a line segment.
\item  If the hyperbolic region of a player is 2-dimensional and his reaction time is different from the reaction times of all other players,
          then his hyperbolic region is non-convex. 
 \item   Every non-convex hyperbolic region is simply connected and star-shaped (as seen by the player who controls the region).
\item The borders between regions are either hyperbolic arcs (when the reaction times of the players owning the regions are unequal) 
          or straight segments (when the  reaction times of the players are equal). 
\end{enumerate}

\subsection*{\normalsize General case}
We can now understand the problem of drawing dominance area in full generality. Using the slowness for the players,  $k_1, k_2$, the locus of points reached at the same time is given by
\begin{equation}
      k_1 \, r_1 - k_2 \, r_2 = t_0 .
\label{eq:CO6}
\end{equation}
Assume that we define new radial coordinates $\tau_i = r_i / A_i$ which have dimensions of time. These coordinates parametrize the points of the plane in terms of the time needed for each player to reach each point. 
%In analogy with radial distance $r$, 
For this reason, we may call the variable $\tau_i$ the \textbf{personal time} for the player P$_i$.  In terms of the personal times, the locus of points which can be reached simultaneously by both players is still a single-branch hyperbola:
$$
        \tau_1 - \tau_2 =   t_0 .
$$
However for practical applications, such as soccer, the transformed variables are not as convenient as they would be in abstract mathematics. An understanding of the locus based on the standard coordinates is necessary.

It turns out that Rene Descartes, in his masterpiece \textsf{La G\'eom\'etrie}\cite{Descartes,Descartes2}, in which he introduced the new area of \textit{Analytical Geometry}, wrote a section on \textit{generalized conics} which he called \textit{bifocal ellipses}. According to his definition, a bifocal ellipse is the locus of points whose distances $r_1, r_2$ from two other points (the foci) satisfy the equation
\begin{equation}
     r_1 + \lambda \, r_2 = d ,
\label{eq:CO1}
\end{equation}
where $\lambda$ and $d$ are two real constants and $\lambda\ne0$. Descartes introduced these curves motivated by an optics problem. As light rays pass from air to glass, they refract. He wanted to determine the unique point in the glass where the refracted rays meet. Today, the curves defined by equations of the form  \eqref{eq:CO1} are known as \textbf{Cartesian ovals}.

The Cartesian oval \eqref{eq:CO1} is not empty if and only if
$d \ge 2c\, \max\{1,-\lambda\}$ when $1+\lambda>0$ and  $d\le2c\,\min\{1, -\lambda\}$ when $1+\lambda<0$, where $2c>0$ is the distance between the two foci.  
I will explain how this condition is established  in the following.  In our case, 
$$
     \lambda=-{A_1\over A_2}, \quad d= A_1\, t_0 .
$$
Without loss of generality, I will assume that the first player is the faster one. However, we cannot necessarily require that he will also have the better reaction time. Therefore
\begin{equation*}
           \lambda < -1 ,
\end{equation*}
and $d$ can have either sign or vanish. The significance of $d$ is that it parametrizes the distance the first player gains/loses due to better/worse response time. The limiting cases
$\lambda=1$ or $t_0=0$ or both have already been discussed.  Hence, the interest in this section is when none of these conditions are valid; that is,
$\lambda\ne1$ and $d\ne0$. In fact, these conventions are standard for the Cartesian ovals.
If the Cartesian oval is empty, there is no point on the plane that the slower player can reach simultaneously with the faster player. Hence the entire plane is dominated by the faster player. When the Cartesian oval is non-empty, then there is a region of the plane which P$_2$ dominates. This happens when,
\begin{equation}
       d \le 2c    ~\Leftrightarrow~  t_2  \le  t_1+{2c\over A_1} .
 \label{eq:CO3}
\end{equation}
This sets an upper bound for the reaction time of the slower player.
The quicker he reacts compared to the faster player, the greater the area he dominates. (See the figure on page \pageref{fig:CDs}.) The slower he reacts, the smaller the area he dominates. And his
window of opportunity completely disappears if he reacts by $2c/A_1$ after the fast player. 
Such a delay is exactly the time the fast player needs to reach the slower player. From there to any point, it is always a winning race for the fast player.
 %Hence, if the slow player is also slow in his reactions, the fast player comes to his point and from there he can go first in any other point in the plane. 
%In particular, the advantage of his better reaction time must be at least enough to allow him to cover the distance which separates him from the faster player.
For the reason just discussed, I will call equation \eqref{eq:CO3},  the \textbf{compensating condition} and from now on I will assume that it is valid since, otherwise, the  fast player dominates the entire plane. Also, for obvious reasons, I will name the dominance area of a player the \textbf{Cartesian region} and the resulting diagram the \textbf{Cartesian diagram} (which should not be confused with the concept of a Cartesian coordinate system).

We can adopt the same coordinate system we used previously and do some algebraic manipulations once more to rewrite  equation  \eqref{eq:CO6} 
in an implicit functional form $f(x,y)=0$:
$$
   \left\lbrack  (k_1^2-k_2^2) (x^2+y^2+c^2)-2c(k_1^2+k_2^2)\, x-t_0^2 \right\rbrack^2 = 4t_0^2k_2^2\, \left\lbrack (c^2+x^2)^2+y^2\right\rbrack .
$$
If we set
$$
      D=k_1^2-k_2^2, \quad S=k_1^2+k_2^2,
$$
we find the slightly simpler form
$$
   \left\lbrack  D (x^2+y^2)]-2cS\, x+Dc^2-t_0^2 \right\rbrack^2 = 2t_0^2(S-D)\, \left\lbrack (c^2+x^2)^2+y^2\right\rbrack .
$$
In general, the boundary between the players is not a necessarily a straight line, circle or hyperbola anymore. Since the defining equation is a quartic one, the actual curve is more complicated. However, the problem to determine its shape and properties is not hard as it may appear at first sight. 

It is more convenient to use polar coordinates $r, \phi$ to write the equation of the boundary. And since the defining equation already contains the distances from the two players, it is simpler to use one of the players as the origin --- say, the slower player P$_2$. We will also measure the polar angle $\phi$ from the line that joins the two players.  Since $r=r_2$, we have $r_1=-\lambda r+d$, from which
$$
    r_1^2 = d^2+\lambda^2r^2-2\lambda d \, r .
$$ 
The quantity $r_1^2$ can be eliminated from this equation by using the law of cosines in the triangle  $\triangle\text{PP}_1\text{P}_2$:
$$
    r_1^2 = r^2+4c^2+4cr\, \cos\phi.
$$
After the substitution and a simple rearrangement of terms,
\begin{equation}
   (1-\lambda^2)\, r^2 + 2(2c\cos\phi+\lambda d)\, r + 4c^2-d^2 = 0.
\label{eq:CO2}
\end{equation}
This equation determines the function $r=r(\phi)$. Since it is a quadratic equation, it may have two, one or no solutions. Because of the compensating equation, it has at least one solution. Therefore, its discriminant $D\ge0$. But it cannot be $D=0$ since the coefficient of the $r$ contains the polar angle.
That is, $D$ is a function of $r$ and even if it vanishes at a value of $\phi$, it cannot vanish at all angles.
So, in general, there must be two solutions of the equation, say $\rho_1(\phi)$ and $\rho_2(\phi)$. 
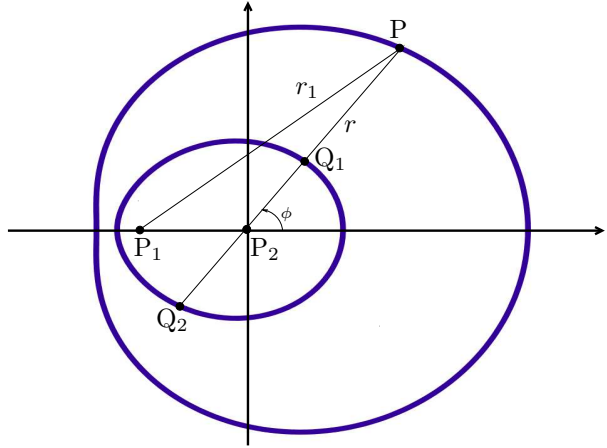
\begin{figure}[h!]
\centering
\setlength{\unitlength}{1mm}
\begin{picture}(80,60)
\put(0,0){\includegraphics[width=8cm]{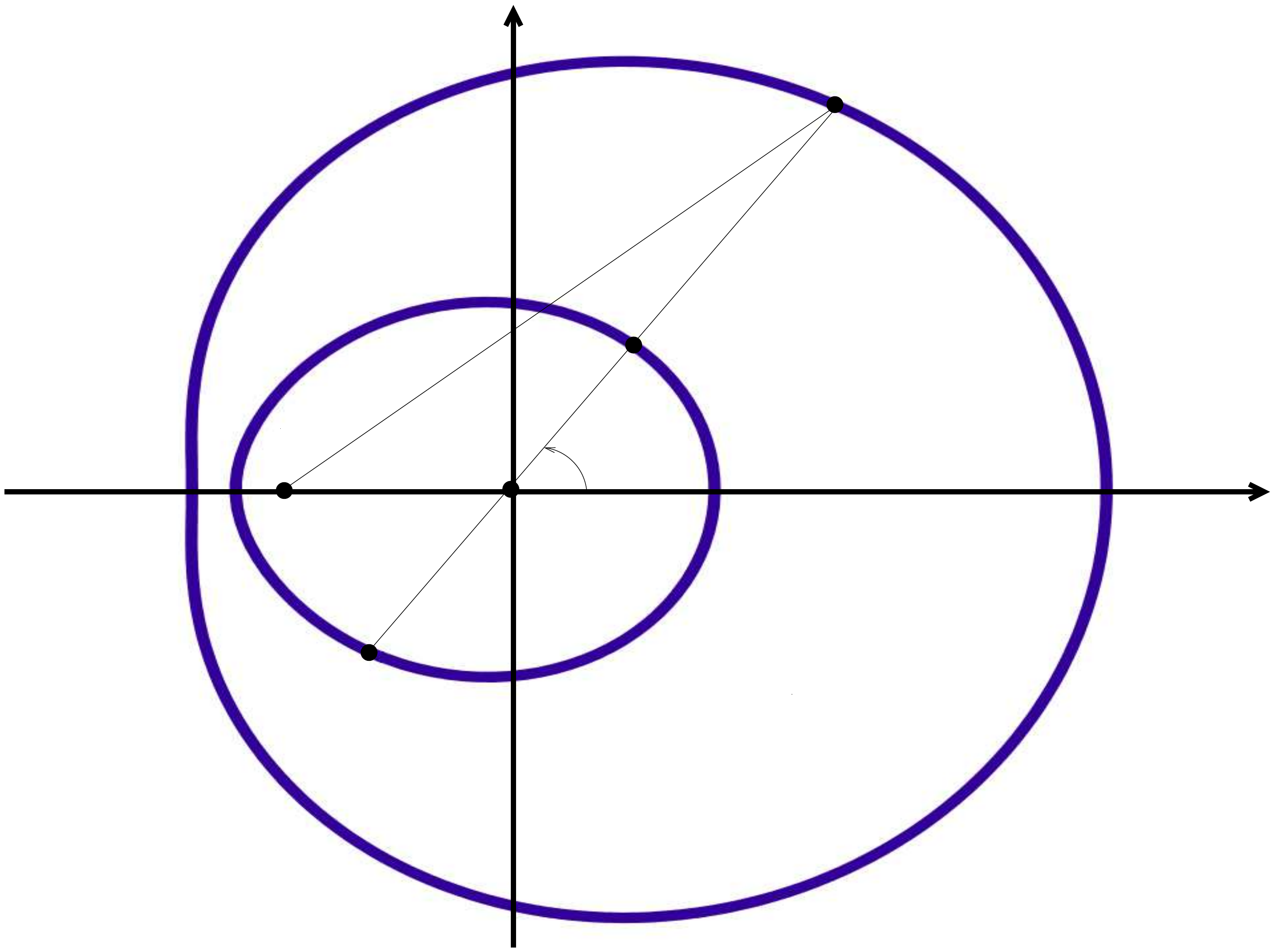}}
\put(17,25.5){\footnotesize P$_1$}
\put(32.5,25.5){\footnotesize P$_2$}
\put(45,42){\footnotesize $r$}
\put(38.5,47){\footnotesize $r_1$}
\put(36.5,30.5){\tiny $\phi$}
\put(51,54.5){\footnotesize P}
\put(41,37.5){\footnotesize Q$_1$}
\put(20,16){\footnotesize Q$_2$}
\end{picture}
\caption{\footnotesize The Cartesian oval.}
\label{fig:oval}
\end{figure}
In other words, Descartes oval is a closed curve with two branches, $r=\rho_1(\phi)$ and $r=\rho_2(\phi)$. However, there is a little more to understand about
this result.  With Figure \ref{fig:oval} in mind, say that $\rho_1(\phi)$ is the solution which corresponds to  the length of P$_2$P. The second solution $\rho_2(\phi)$ sometimes corresponds to  the length of the  segment P$_2$Q$_1$ and sometimes to the segment P$_2$Q$_2$ which is pointing on the opposite ray.   This might sound strange but observe that, as $\phi$ performs by a full cycle, its cosine does not maintain the same sign. In the first and fourth quadrant, $\phi\in(-\pi/2,~\pi/2)$, $\cos\phi$ is positive and in the second and  third quadrants, $\phi\mapsto\phi+\pi$, $\phi\in(-\pi/2,~\pi/2)$,
 the cosine is negative. For those values that the cosine is negative, let's write equation \eqref{eq:CO2},
$$
   (1-\lambda^2)\, r^2 + 2(2c\cos(\phi+\pi)+\lambda d)\, r + 4c^2-d^2 = 0,
$$
in the form  
$$
   (1-\lambda^2)\, (-r)^2 + 2(2c\cos\phi-\lambda d)\, (-r) + 4c^2-d^2 = 0.
$$
We clearly now see that this gives a solution for the opposite value of $\lambda$ at the polar value $\phi$, but along the opposite ray. 

It is immediate from the previous discussion that each branch of the Descartes oval corresponds to a different sign of $\lambda$. Since we are are looking 
at that branch which relates to a  difference of the lengths, intuitively we understand that it is the larger oval that we are looking for. 
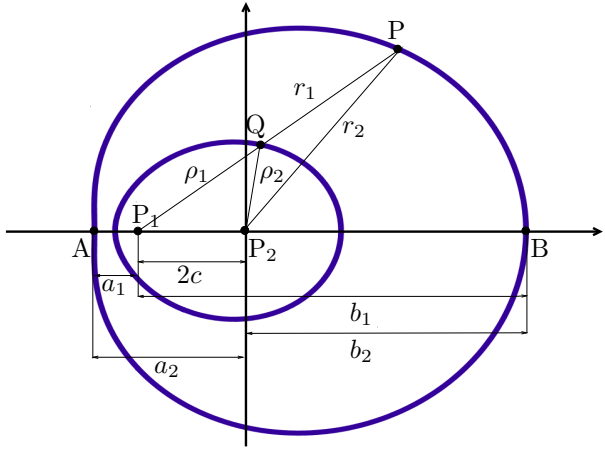
\begin{figure}[h!]
\centering
\setlength{\unitlength}{1mm}
\begin{picture}(80,60)
\put(0,0){\includegraphics[width=8cm]{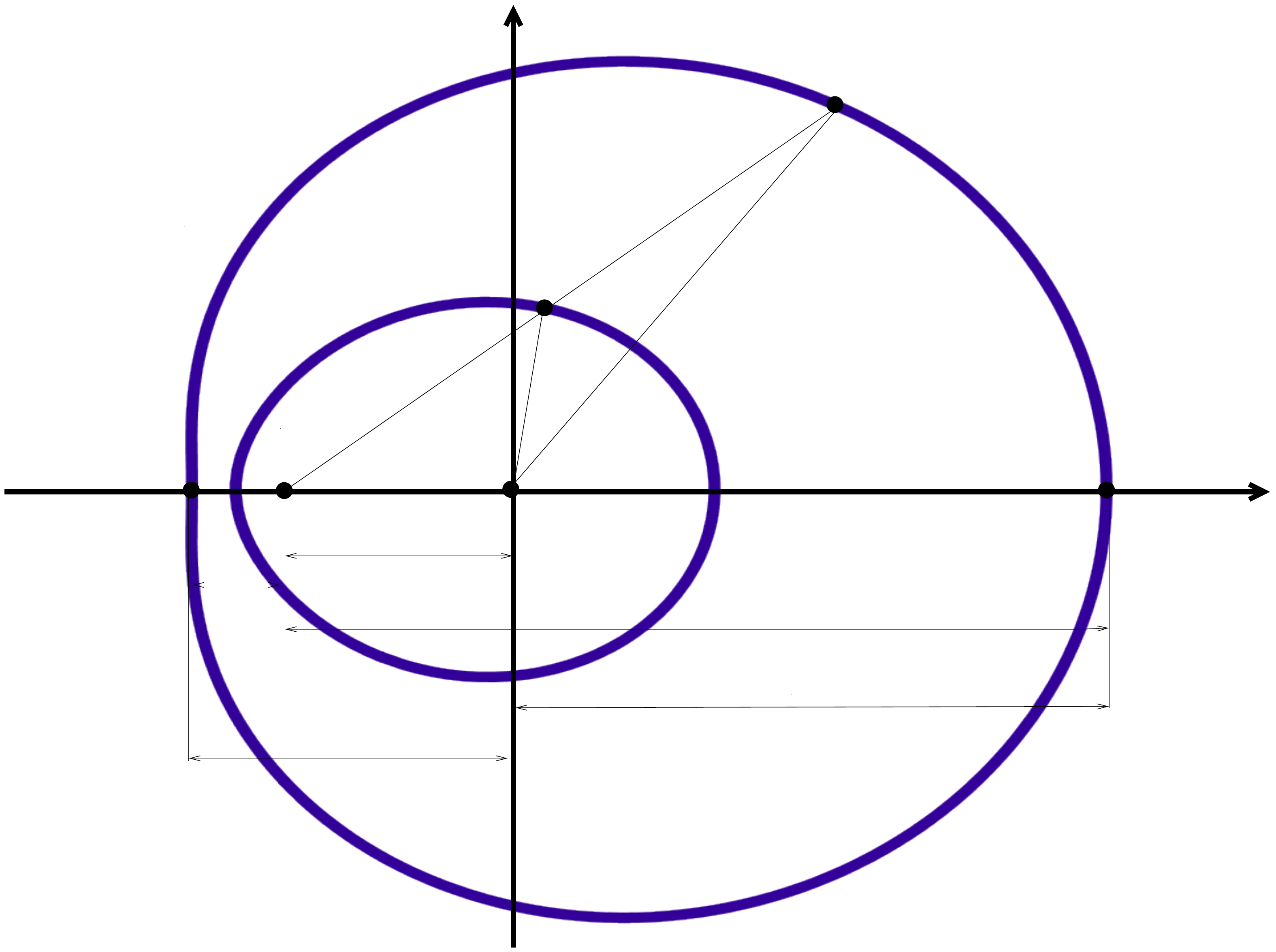}}
\put(17,30){\footnotesize P$_1$}
\put(32.5,25.5){\footnotesize P$_2$}
\put(45,42){\footnotesize $r_2$}
\put(38.5,47){\footnotesize $r_1$}
\put(34,36){\footnotesize $\rho_2$}
\put(24,36){\footnotesize $\rho_1$}
\put(51,54.5){\footnotesize P}
\put(32.1,42){\footnotesize Q}
\put(9,25.5){\footnotesize A}
\put(70,25.5){\footnotesize B}
\put(20,10){\footnotesize $a_2$}
\put(46,12){\footnotesize $b_2$}
\put(13,21){\footnotesize $a_1$}
\put(46,17){\footnotesize $b_1$}
\put(23,22){\footnotesize $2c$}
\end{picture}
\caption{\footnotesize Identification of the branches in the Cartesian oval. The outer branch corresponds to $r_1-|\lambda| r_2=d$ and the inner branch 
               $r_1+|\lambda| r_2=d$. Hence the Cartesian region of the slower player is the area inside the outer branch and the Cartesian region of the 
               faster   player is its complement in the plane.}
\label{fig:oval3}
\end{figure}
However, let's prove it formally. Let Q and P be two distinct points on the two branches as shown in the Figure \ref{fig:oval3} with distances from the two players $\rho_1, \rho_2$ and $r_1, r_2$ respectively. If Q was satisfying
$$
      \rho_1 - |\lambda| \, \rho_2 = d,
$$
we will show that it is impossible for P to satisfy 
$$
      r_1 + |\lambda| \, r_2 = d .
$$
Indeed, subtracting the two relations gives
$$
      \rho_1 - r_1 =   |\lambda| \, (r_2  + \rho_2 ) ,
$$
which is impossible since the two sides have opposite signs given that $\rho_1< r_1$. 

Now let's return to the compensating condition and discuss how it emerges from the oval. Consider the extreme point B.
Let the distances of B from the players be $(b_1,b_2)$. Then
$$
       b_1 - |\lambda|\, b_2 = d.
$$
Also, from the figure, it is easily established that
$$
     b_1-b_2=2c.
$$
From the above equations,  a quick calculation gives:
\begin{align*}
     b_1= {d-2c\,|\lambda| \over 1-|\lambda| }, \quad b_2={d-2c \over 1-|\lambda|} .
\end{align*}
Since the denominators are negative, the numerators must be non-positive. Hence $d\le\min\{2c, 2c|\lambda|\}=2c$.
\begin{figure*}[p]
\vspace{-3cm}
\rotatebox{90}{%
 \begin{tabular}{ccc}
 \multicolumn{3}{ p{22cm} }{ 
 \stepcounter{figure} 
 Figure \thefigure: \label{fig:CDs} 
   \footnotesize This figures show the dominance areas for two players separated by a distance $2c=4$. The slower player is located at the origin and his dominance area is that which is enclosed  by the Cartesian oval.
Each diagram of the figure contains three curves which correspond to the boundary of the dominance areas for the three ratio of the speeds $\lambda=1.1, 1.2, 1.3$ at the same time delay difference $t_0$. As we move from the first (top left) to the last (bottom right) diagram, the delay difference which starts advantageous for the slower player decreases until it reverses and becomes advantageous for the faster player. 
In particular, $d=-7, -6, -4, -2, 0, 3$. Notice how the Cartesian ovals degenerated to Apollonius circles when the time delay vanished. At $d=4$, the dominance areas disappear.} \\
\includegraphics[height=7cm]{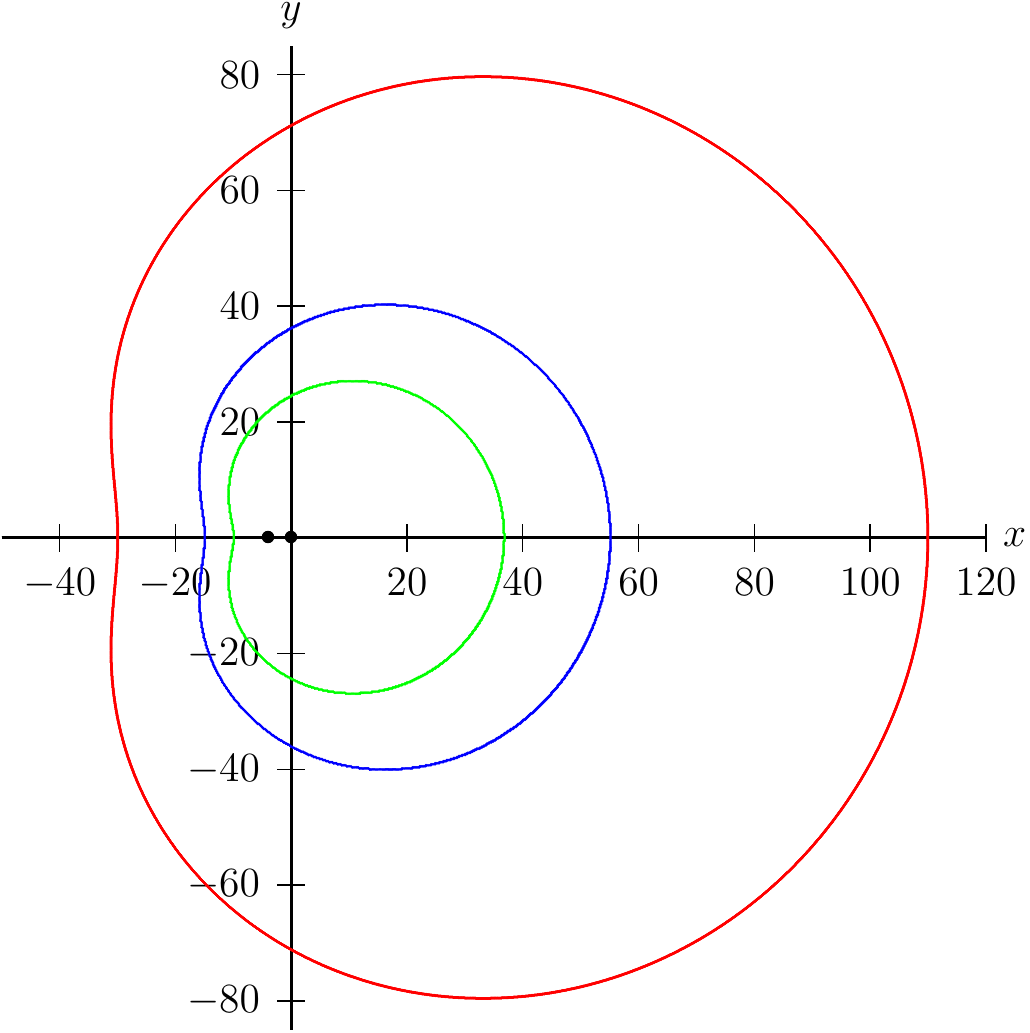}   & \includegraphics[height=7cm]{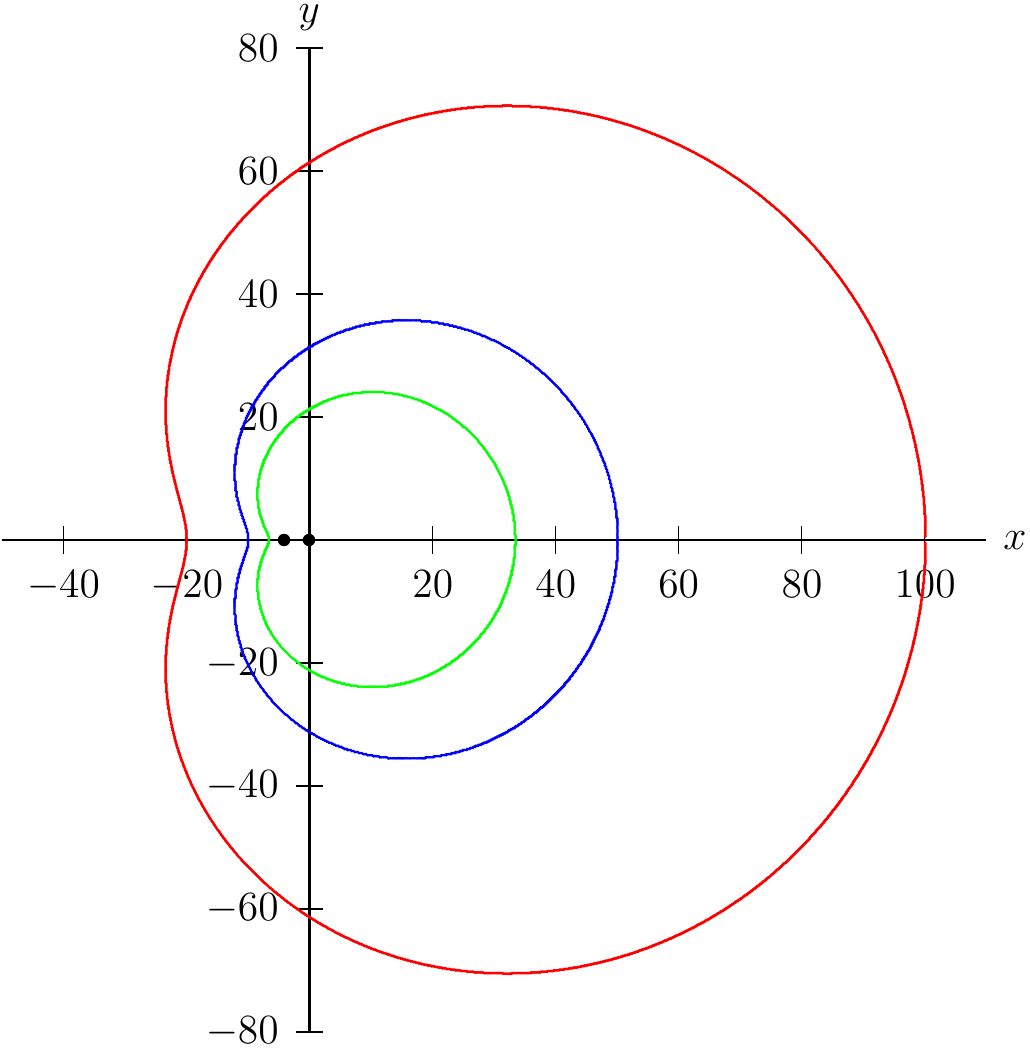}  & \includegraphics[height=7cm]{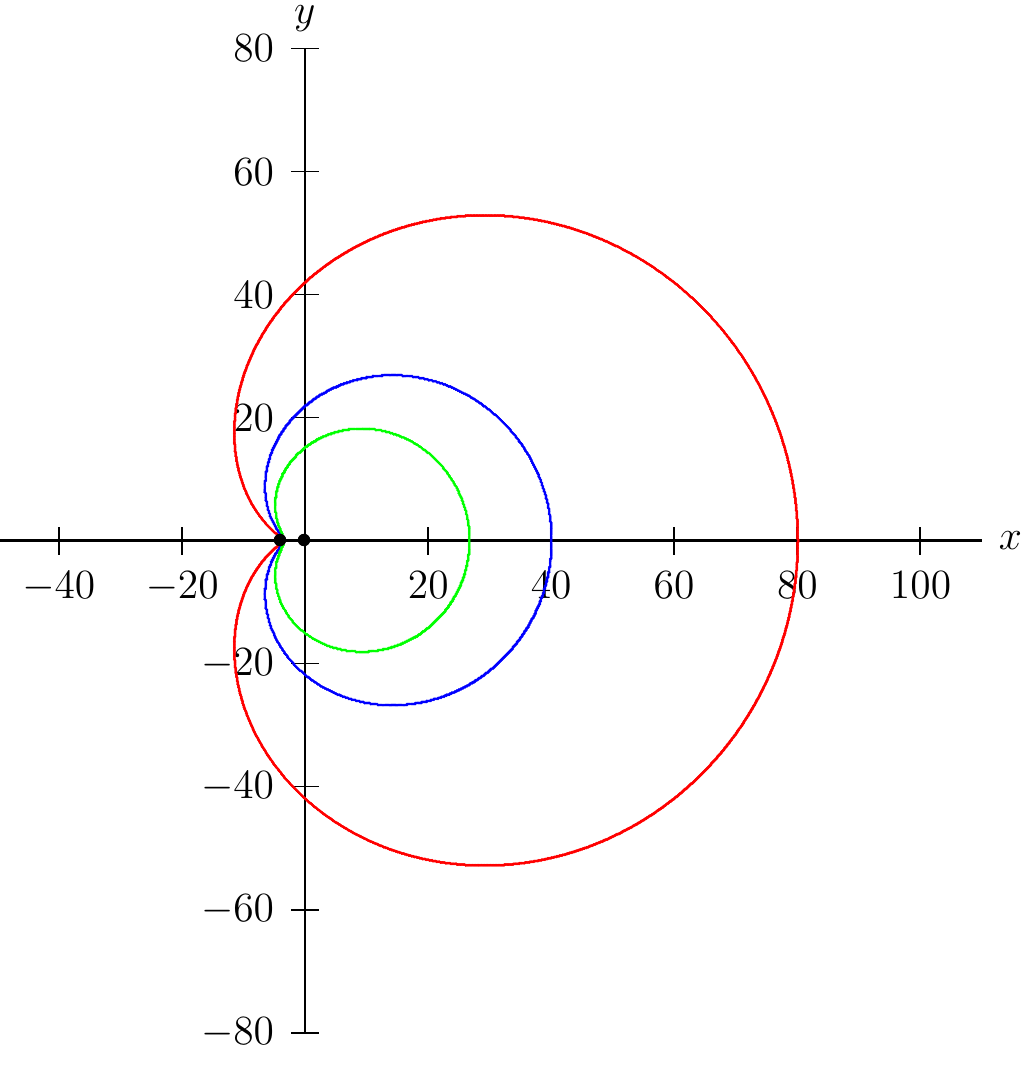}    \\ \includegraphics[height=7cm]{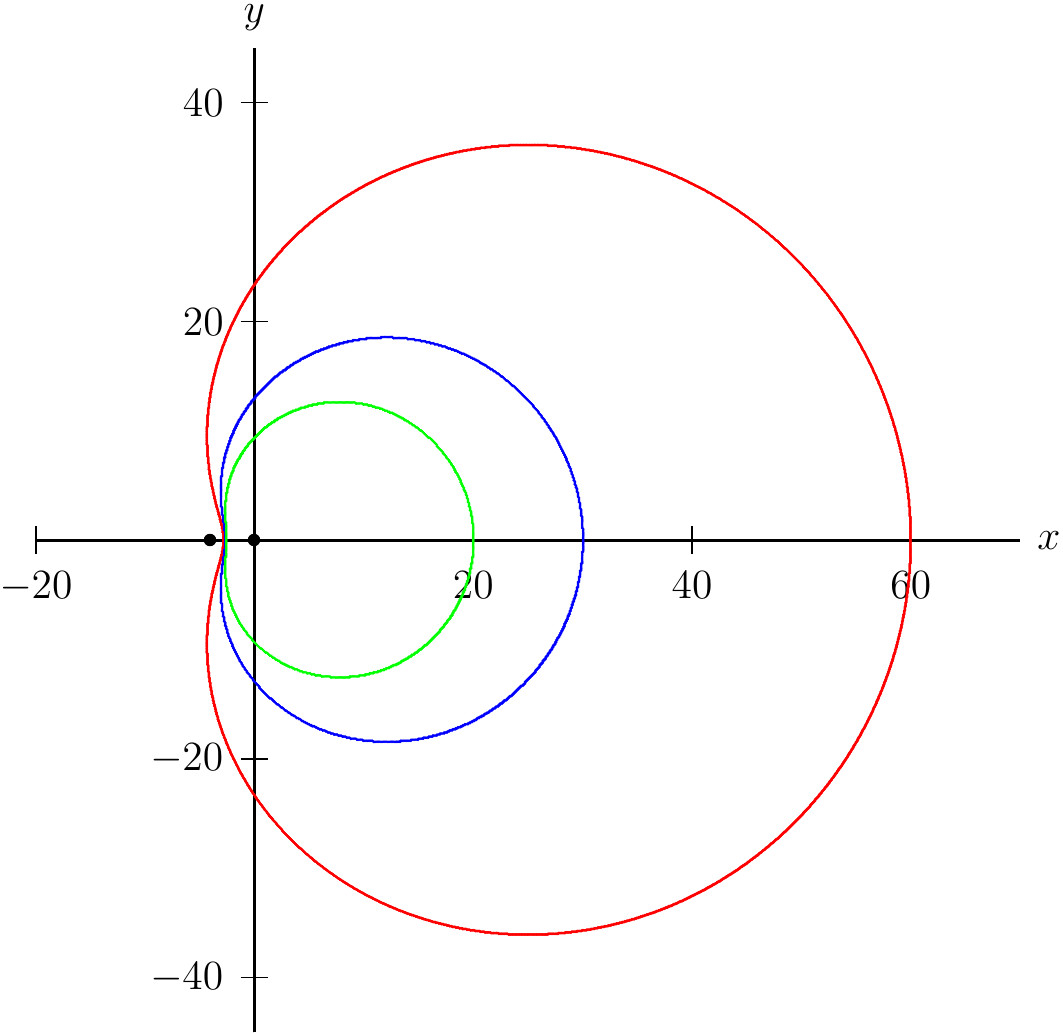}    & \includegraphics[height=7cm]{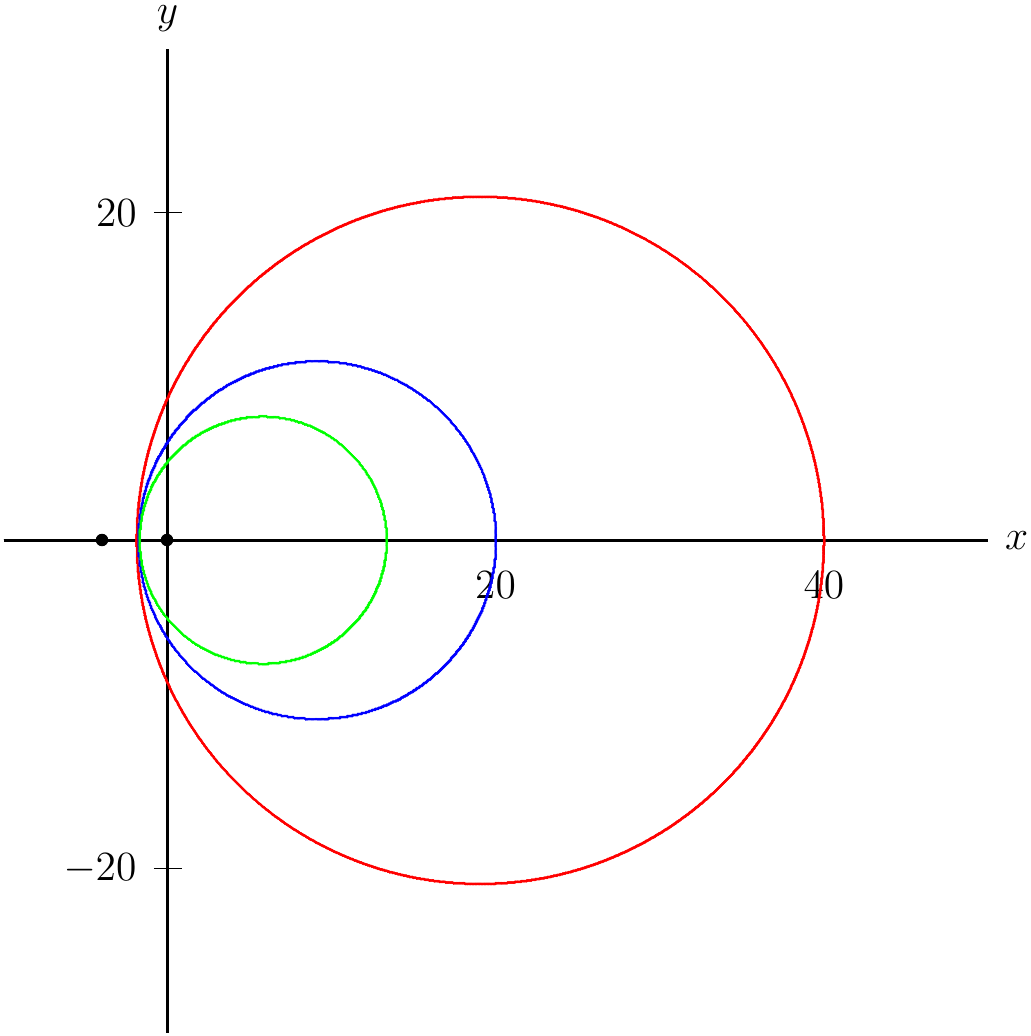}  & \includegraphics[height=7cm]{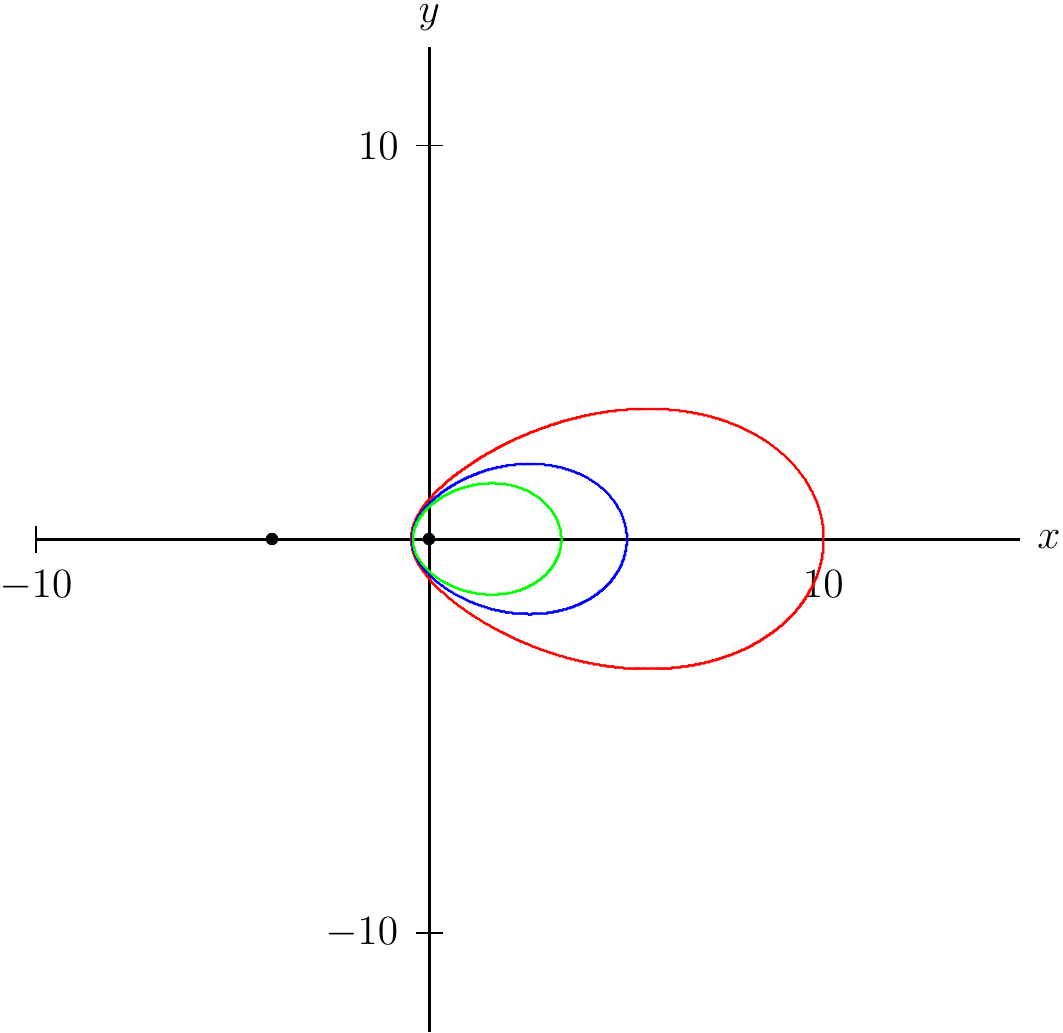}      
 \end{tabular}        
                  }%end_rotate
\end{figure*}
 \let\foo\thefigure

Let's do the same calculation with point A.
Let the distances of A from the players be $(a_1,a_2)$. Then
$$
       a_1 - |\lambda| \, a_2 = d,
$$
and
$$
    a_2-a_1=2c.
$$
From the four equations,  a quick calculation gives:
\begin{align*}
     a_1= {d+2c\,|\lambda| \over 1-|\lambda| }, \quad a_2={d+2c \over 1-|\lambda|} .
\end{align*}
Again, since the denominators are negative, the numerators must be  non-positive. However, this is possible only for $d\le -2c|\lambda|$.
There is a discrepancy in the region $-2c|\lambda| < d < 2c$. This shows that the calculation for point A is wrong in this region. Indeed this happens since 
the calculation assumed that both foci are inside the oval. However, sometimes this is not the case. When P$_1$ is outside the dominance area of P$_2$,
we have 
$$
      a_1+a_2=2c.
$$
Then
\begin{align*}
     a_1= {d+2c\,|\lambda| \over 1+|\lambda| }, \quad a_2={2c-d \over 1+|\lambda|} .
\end{align*}
Figure \foo\  shows the  Cartesian region of player P$_2$ for various values of the parameters $\lambda$ and $d$. It can be clearly 
observed that in some cases player P$_1$ is inside this region and in some cases he is outside of the region in accordance with the previous results.

Notice the  fact (already known to coaches and players) that has been reinforced by our calculation.
A slow defender P$_2$ with fast reactions,
$$
    t_2 \le t_1 - {2c\over A_2} ,
$$
 can actually neutralize a fast attacker since his dominance area contains the fast player. Only if the ball is played outside the dominance area of P$_2$, the fast player will intercept 
 it first. In none of the previous cases, the dominant region of a player contained the other player (except in the extreme limit in which a player was controlling only a single point). 
 % Although he does not control the entire plane, in an actual game a pass is sent to the vicinity of the players. If the defender makes sure that he positions himself such that his dominance area includes the attacker, he has has good chances to neutralize him. This is ultimately related to a fast reaction time. Without incorporating reaction time, the dominance region of a player cannot include another player.

%%%%%%%%%%%%%%%%%%%%%%%%%%%%%%%%%%%%%%%%%%%%%%%%%%%%%%%%%%%%%%%
\section{A Simple Example: 2 vs 2}

Just to demonstrate the diagrams discussed in this work, let's imagine a 2 vs 2 game between four players with characteristic speeds and time delays as given in Table
\ref{table:2v2}.
\begin{table}[h!]
\centering
\begin{tabular}{| c | c | c | c| } \hline\hline
\textbf{player} & \multicolumn{2}{c|}{\bf characteristic speed} & {\bf time delay} \\ \hline 
                       &   mi/h & m/s  & s \\ \hline \hline
Player 1  & 30 & 13.33 & 0 \\ \hline
Player 2  & 30 & 13.33 & 0.5 \\ \hline
Player 3 & 40 & 17.78 & 0.5 \\ \hline
Player 4  & 35 & 15.56 & 0 \\ \hline
\end{tabular}
\caption{Characteristic speeds and time delays of  four players involved in a 2 vs 2 game. The values chosen have been selected for demonstration 
only. A follow up paper is constructing realistic diagrams using real tracking data \cite{EDG}.}
\label{table:2v2}
\end{table}
The pitch will be assumed  to have dimensions $40\times25$ in meters.  Players P$_1$ and P$_3$ make one team and the other two players the opposing team and at some moment of time the players are located at the points  P$_1(5,5)$, P$_2(10,5)$, P$_3(10,20)$, P$_4(15,20)$. We will assume that all players have zero speeds at this moment.

\begin{figure}[ht!]
\centering
\setlength{\unitlength}{1mm}
\begin{picture}(100,60)
\put(0,0){\includegraphics[width=10cm]{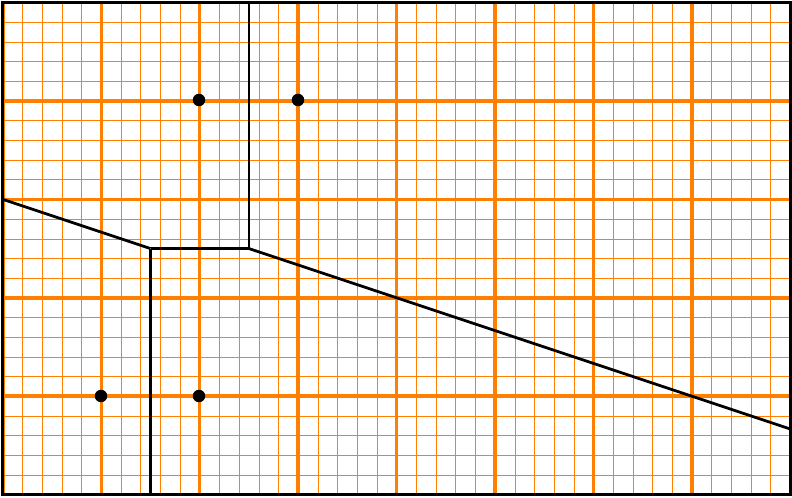}}
\put(8.5,14){\footnotesize P$_1$}
\put(26,14){\footnotesize P$_2$}
\put(20.5,51.5){\footnotesize P$_3$}
\put(38,51){\footnotesize P$_4$}
\end{picture}
\caption{\footnotesize The standard Voronoi diagram of the 2v2 game. This and the following diagrams in this section have been drawn with a grid on the background such that the reader can do approximate back-of-the-envelop calculations fast and draw conclusions easily (such us compare areas).}
\label{fig:SD}
\end{figure}

The standard Voronoi diagram can be drawn very quickly by drawing the perpendicular bisectors to the segments that join any two players. The resulting diagram is seen
in Figure \ref{fig:SD}.
This, as we have discussed, corresponds to the case that the data of Table \ref{table:2v2} are ignored and we adopt equal characteristic speeds and equal delays
for all four players.

Then we draw the Apollonius diagram assuming that all time delays  are the same.
\begin{figure}[hb!]
\centering
\setlength{\unitlength}{1mm}
\begin{picture}(100,60)
\put(8.5,14){\footnotesize P$_1$}
\put(26,14){\footnotesize P$_2$}
\put(20.5,51.5){\footnotesize P$_3$}
\put(38,51){\footnotesize P$_4$}
\put(0,0){\includegraphics[width=10cm]{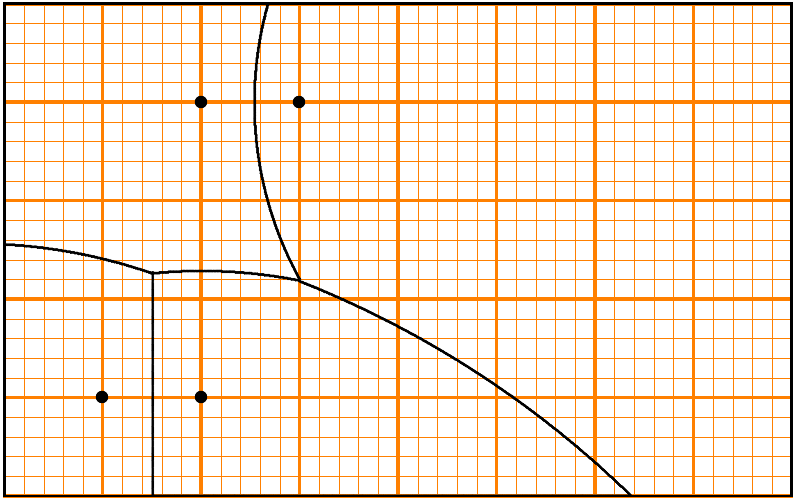}}
\end{picture}
\caption{\footnotesize The Apollonius  diagram of the 2v2 game. Compared to the standard Voronoi diagram, the fastest player P$_3$ has expanded his dominance region by acquiring space from all remaining players. The second fastest player has also expanded his region by taking area from player P$_2$.
However, the boundary between players P$_1$ and P$_2$ who have equal speeds has  remained unchanged; none of them can claim area from the other player's region.
}
\label{fig:2v2Apollonius}
\end{figure}
Besides the expected result that the faster the player, the more his dominance region will increase relative to the standard diagram, notice a nice additional fact which emerges from the Apollonius diagram: When speeds are taken into account, the diagram shows clearly that  a fast player (player P$_3$ here)  can go through the opposing team easier. In particular,  he can intercept  a through ball from his teammate behind the defending line  (the line P$_2$P$_4$ here)  without the defense (players P$_2$ and P$_4$ here) having the opportunity 
to stop him. This is consistent with our experience.

Since, only the ratio of speeds enters the calculations for the Apollonius diagram, the diagram would be identical if we were to replace the characteristic speeds $V_1, V_2, V_3, V_4$ by $\mu V_1, \mu V_2, \mu V_3, \mu V_4$, where $\mu$ any positive real number. This might appear counterintuitive since doubling, for example, the speeds, it also
doubles the difference in speeds among players.

\begin{figure}[h!]
\centering
\setlength{\unitlength}{1mm}
\begin{picture}(100,60)
\put(8.5,14){\footnotesize P$_1$}
\put(26,14){\footnotesize P$_2$}
\put(20.5,51.5){\footnotesize P$_3$}
\put(38,51){\footnotesize P$_4$}
\put(0,0){\includegraphics[width=10cm]{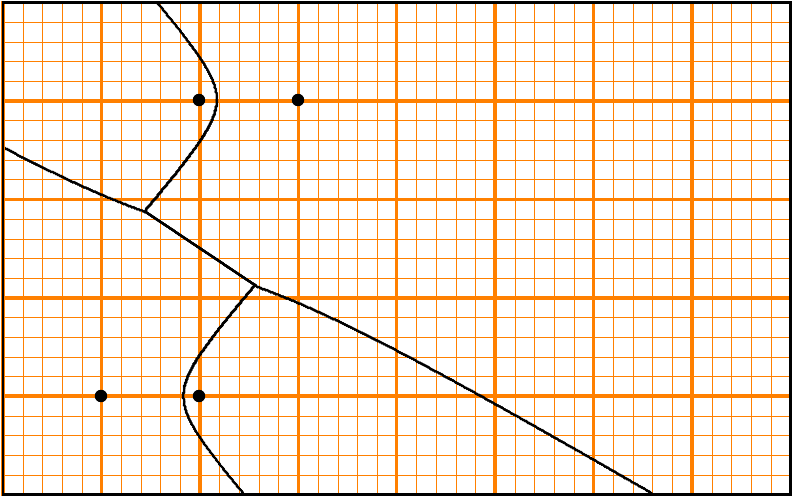}}
\end{picture}
\caption{\footnotesize  The  hyperbolic  diagram for  the 2v2 game. Compared to the standard diagram, players P$_1$ and P$_4$ who have faster reactions have gained areas at the expense of the other two players.}
\label{fig:2v2Hyperbolic}
\end{figure}

The third case in which all players are assumed to have the same characteristic speeds (to be assumed 30 mi/h) but different delay times is drawn in Figure \ref{fig:2v2Hyperbolic}. 
For the hyperbolic diagram, since only the differences of  the time delays  enter  the calculations, this diagram would be identical  if we were to replace the any time delays $t_1, t_2, t_3, t_4$ by $t_1+T, t_2+T, t_3+T, t_4+T$, for any real number $T$. 
Hence, we can always set the delay time of a specific player to zero and measure the delay times of the remaining players relative to  that selected player.

\begin{figure}[h!]
\centering
\setlength{\unitlength}{1mm}
\begin{picture}(100,60)
\put(8.5,14){\footnotesize P$_1$}
\put(26,14){\footnotesize P$_2$}
\put(20.5,51.5){\footnotesize P$_3$}
\put(38,51){\footnotesize P$_4$}
\put(0,0){\includegraphics[width=10cm]{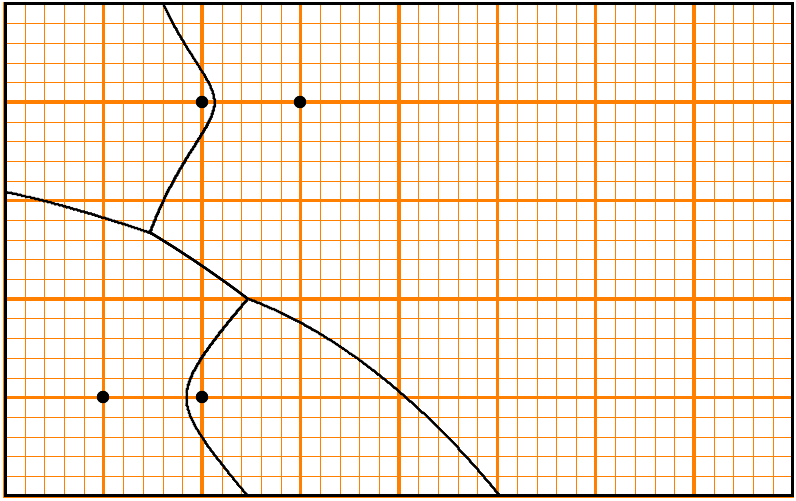}}
\end{picture}
\caption{\footnotesize  The  Cartesian  diagram for the 2v2 game. Compared to the standard Voronoi diagram, each player gains or loses area depending on his characteristic speed and delay time relative to the rest players. }
\label{fig:2v2Cartesian}
\end{figure}

Finally, Figure \ref{fig:2v2Cartesian} presents the diagram when both sets of parameters, the characteristic speeds and the time delays, are taken 
into account.
This diagram remains the same if all time delays are shifted by $T$ (as in the hyperbolic diagram) but
it does not remain the same if the characteristic speeds are scaled by $\mu$ since the quantity $d$ in equation \eqref{eq:CO1} depends on a single speed and not on a ratio
of speeds. 

Of course, the actual diagrams are affected greatly by the choice of the values for the characteristic speeds and time delays. The reader is encouraged to draw his own diagrams for various ranges of these parameters in order to get a feeling of the effect.

%%%%%%%%%%%%%%%%%%%%%%%%%%%%%%%%%%%%%%%%%%%%%%%%%%%%
\section{Discussion and Conclusion}

In this article, I have argued that the standard Voronoi diagrams that are widely used for computing dominance area in soccer are inadequate to provide
an exact measure of the concept intended to communicate. The fundamental inadequacy  stems from the fact that they are built to compute proximity area for still objects and do not include parameters related to the motion of moving objects. To understand the correct concept that should replace the standard Voronoi diagrams, a careful  examination of the underlying dynamics must  be made. As I have stated,  by dynamics I imply the mechanism at work for the concept of `dominance area' and not necessarily the real fundamental dynamics  in the sense of theoretical physics. For this reason, I have proposed the term `soccerdynamics' not only as an analogy but more as a distinct name that clearly separates soccer modeling using physical concepts from other physics areas that involve the term `dynamics' (such as electrodynamics).  For the dominance area, when the mechanism is studied using simple kinematics, one can immediately realize that the standard Voronoi diagrams must be replaced by new diagrams whose regions are not necessarily convex polygons anymore and they exhibit some new properties.

We have previously mentioned that  mathematicians have already written down one of the variations of the Voronoi diagram which emerged from our study. In particular, our Apollonius diagram is  known as \textbf{multiplicatively weighted Voronoi diagram} and uses the modified metric:
$$
       \tilde d(\text{P}, \text{P}_i) =  {d(\text{P}, \text{P}_i) \over W_i} .
$$
This statement also applies for the other variations of the Voronoi diagrams  we discovered.  Our hyperbolic digram is  known to mathematicians as \textbf{additively weighted Voronoi diagram} and uses the modified metric:
$$
       \tilde d(\text{P}, \text{P}_i) =  d(\text{P}, \text{P}_i) - w_i ;
$$
and our Cartesian digram is  known as \textbf{compoundely weighted Voronoi diagram} and uses the modified metric:
$$
       \tilde d(\text{P}, \text{P}_i) =  {d(\text{P}, \text{P}_i) -w_i\over W_i} .
$$
In the above equations, $d$ is the standard Euclidean metric and the numbers $w_i, W_i$ are called the \textbf{additive} and \textbf{multiplicative weights} respectively of the point P$_i$. A priori they have nothing to do with speeds and delay  times; they are just free parameters used by mathematicians to establish their theories. They receive appropriate meaning within the setting where they are applied.  Unfortunately the literature of these modified distances is  not as extensive as the literature of the standard Voronoi diagram. This is especially true
for the compoundely weighted Voronoi diagram. I have not encountered any discussion or even reference to the Cartesian oval.  The additively and multiplicative weighted diagrams have been rediscovered from time to time in the research of very distinct areas of science. These original papers have been reported in \cite{OkabeEtAl,AKL}.  The three straightforward variations of the standard Voronoi  diagram (which we also discovered) have  served as motivation for mathematicians to propose even more complicated modifications of the  Votronoi diagram; today the list of possible variations appears to be very long. One  has only to look at the contents of  the previously cited books to verify this claim.

I have already discussed some of the assumptions I have made in order to advance this study. However, it is worthwhile to add some additional comments.
Returning to the vectorial description of the accelerated motion of the players before the simplifying assumption to convert the equations
to algebraic ones, we can easily derive
\begin{equation}
    \vec r_i =  \vec A_i \, (t-t_i), \quad i=1,2,
 \label{eq:55}
\end{equation}
where
$$
    \vec A_i = { \vec v_i + \vec V_i \over 2 } , \quad i=1,2,
$$
is the average of the initial and final velocities.  Taking the magnitude of $\vec r_i$ in \eqref{eq:55},
\begin{equation*}
     r_i =   A_i \, (t-t_i), \quad i=1,2.
\end{equation*}
Although this result looks identical to the the algebraic equation
\eqref{eq:4}, it is not. By squaring the defining equation of $\vec A_i$,
$$
    4\vec A_i^2 =  \vec v_i ^2+ \vec V_i^2 +  2 \vec v_i \cdot \vec V_i  , \quad i=1,2,
$$
or 
$$
    4 A_i^2 =  v_i ^2+ V_i^2 +  2  v_i \,  V_i \cos\theta_i , \quad i=1,2,
$$
where $\theta_i$ is the angle formed by $\vec v_i$ and $\vec V_i$. When the initial velocity is off the direction of $\vec r_i$, the player must accelerate
in a way which will orient the final velocity $\vec V_i$ such that the vector $\vec A_i$ will point along $\vec r_i$. Even if $\vec V_i$ has constant magnitude, it does not have constant direction. Hence, $\vec A_i$ has variable magnitude. The locus of points P reached simultaneously by the two players are still the solutions of
$$
   k_1 r_1-k_2 r_2 = t_0 ,
$$
with $t_0=t_2-t_1$, but the players' slownesses, 
$$
  k_1={2\over \sqrt{v_1 ^2+ V_1^2 +  2  v_1 \,  V_1 \cos\theta_1} }, \quad
  k_2={2\over \sqrt{v_2 ^2+ V_2^2 +  2  v_2\,  V_2 \cos\theta_2}},
$$
are now more complicated functions of the point P. 
%(We can use geometry to express $\theta_i$ in terms of the polar angle $\phi_i$ of P relative to player P$_i$ and the angle $\alpha_i$  giving the direction of the initial velocity $\vec v_i$ but this does not add any new insight.)

Adding to the previous discussion a little more from another direction, I should point out that all the curves that have appeared in this study fall in the category of \textit{algebraic curves}. Without expanding to details, I will simply say that, when Cartesian coordinates $x,y$ are used on the plane, these are curves described by an equation of the form $ f(x,y)=0$, with $f(x,y)$ being a polynomial in $x$ and $y$. Mathematicians have established a lot of  abstract results for such curves and books have written  to organize and disseminate these results to the broader mathematical community. For example, the curious reader can  check out the classic book by Coolidge \cite{Coolidge} and the more recent ones \cite{Miranda,Kunz,Bix}. However, he will immediately realize that the mathematicians' interest focus on ideas and concepts which are very abstract and have very little to do (if anything at all) with the exact tracing of the curves which has been the main goal of this paper.

Returning to the discussion of my assumptions, the treatment of the players as being points should also be mentioned. This is true in the standard Voronoi diagrams, as well as in any  study of dominance regions.  But players are not mathematical points. They have size. They also have body parts that move and this motion can directly affect space control. This introduces an additional layer to the problem that adds serious sophistication and difficulty. True soccerdynamics must take this into account.

Therefore,  an immediate consequence of the current work and the comments of this section  is the importance  for player evaluators and advisers to use Voronoi diagrams with some care. Since the dominance area is a concept that depends greatly on many kinematical and physical attributes of the players, the use of the standard Voronoi diagrams is, at least, problematic. I have presented an improved version  but there is still plenty room for further improvement. Then, the question how a team's dominance area during a game correlates with the outcome of the game remains open. Any step forward along this direction (using soccerdynamics) would be a major feat since no one has ever proposed any model to correlate any of the soccer performance indices (the Voronoi diagram being one)  with game outcome. 

Finally, it is obvious that , although I have been referring to soccer, the ideas presented in this article apply to other ball team sports  where the
standard Voronoi diagrams have been applied. The particular details can easily be adjusted to reflect the corresponding sport.

%%%%%%%%%%%%%%%%%%%%%%%%%%%%%%%%%%%%%%%%%%%%%%%%%%%%%

\end{document}